\documentclass[english,12pt,aps,prd,a4paper,preprintnumbers,  floatfix,nofootinbib,showpacs,superscriptaddress,  notitlepage]{revtex4-1} 

\pdfoutput=1

\RequirePackage[colorlinks=true,urlcolor=red,anchorcolor=red,citecolor=red,filecolor=red,linkcolor=red,menucolor=red,pagecolor=red,linktocpage=true,pdfproducer=medialab,pdfa=true]{hyperref}
\usepackage[usenames,dvipsnames]{color}  %%%%%%% usenames,dvipsnames required to get BrickRed
\usepackage{graphicx}
\usepackage{setspace}
\usepackage{upgreek}
\usepackage{braket}

\usepackage{microtype}
\usepackage{multirow}
\usepackage{tabularx}

\usepackage{bm}% bold math
\usepackage{bbm}% blackboard bold math
\usepackage{mathrsfs}
\usepackage[bbgreekl]{mathbbol}% blackboard bold math

\usepackage{caption}
\usepackage{stackengine}
\usepackage{subcaption}
\usepackage{amsmath}
\usepackage{amssymb}
\usepackage[normalem]{ulem}
\usepackage{xcolor}
\usepackage{graphicx}
\makeatletter
\def\p@subsection{}
\makeatother
\usepackage{xcolor}
\usepackage{float}
\usepackage{placeins} %%for FloatBarrier
\usepackage{array}
\definecolor{darkred}{rgb}{0.6,0,0}

\definecolor{linkcolor}{rgb}{0,0,0.5}

\def\gsim{\raise0.3ex\hbox{$\;>$\kern-0.75em\raise-1.1ex\hbox{$\sim\;$}}}
\def\lsim{\raise0.3ex\hbox{$\;<$\kern-0.75em\raise-1.1ex\hbox{$\sim\;$}}}

\def\beqn#1{\begin{equation}\label{#1}}
\def\eeqn{\end{equation}}

\def\beqa#1{\begin{eqnarray}\label{#1}}
\def\eeqa{\end{eqnarray}}

\usepackage{caption}
\usepackage{subcaption}
\usepackage{adjustbox}
\def\Z2{$\mathcal{Z_2}$}

\usepackage{bbold}
\usepackage{ragged2e}
  \newcommand{\rd}[1]{{\color{red}  #1}}
 
\newcommand {\ignore}[1]{}

\usepackage{mathrsfs}
\def\321{$\mathrm{SU(3) \otimes SU(2) \otimes U(1)}$ }

\definecolor{matplotlibBrown}{rgb}{0.65, 0.16, 0.16}

\def\red{\color{red}{}}

\def\rd#1{\textcolor{red}{#1}}

\usepackage{orcidlink}

\newcommand{\AddrIISERB}{Department of Physics, Indian Institute of Science Education and Research - Bhopal, \\ 
Bhopal Bypass Road, Bhauri, Bhopal 462066, India}

\begin{document}

\title{\textcolor{BrickRed}{Realizing Axial $Z'$ Couplings from Flavored Chiral $U(1)_X$ Gauge Symmetries}}

\title{\textcolor{BrickRed}{Light and Heavy $Z'$ from Flavored Chiral $U(1)_X$ Gauge Symmetries: Purely Axial and Mixed Vector-Axial Couplings}}

\author{Hemant Kumar Prajapati~\orcidlink{0000-0001-5104-9427}}\email{hemant19@iiserb.ac.in}
\affiliation{\AddrIISERB}
\author{Rahul Srivastava~\orcidlink{0000-0001-7023-5727}}\email{rahul@iiserb.ac.in}
\affiliation{\AddrIISERB}

\begin{abstract}
\vspace{0.5cm}

%%%%%%%%%%%%%%%%%%%%%%%%%%%%%%%%%%%%%%%%%%%%%%%%%%%%%%%%%%%%%%%%%%%%%%%%%%%%%%

Model independent phenomenological studies, ranging from neutrino to B-physics, often consider effective interactions involving either purely vector (V), purely axial vector (A), or mixed vector and axial vector (V, A) couplings.
While pure vector $Z'$ interactions can naturally emerge in gauged $U(1)_X$ extensions of the Standard Model, such as  the $B-L$ model, generating other coupling structures from a UV complete theory is highly nontrivial.
To realize such couplings, we propose a new class of flavor specific chiral $U(1)_X$ gauge symmetries.
%%%
Gauge anomaly cancellation is achieved by introducing three right-handed neutrinos charged under the $U(1)_X$ symmetry. 
 We systematically classify anomaly free charge assignments and analyze viable ultraviolet completions with minimal scalar content, requiring no additional fermions beyond the three necessary for anomaly cancellation.
We present several benchmark models illustrating the range of possible charge assignments, under which the quark and lepton flavor structures can differ substantially, leading to distinct phenomenological signatures.
In particular, such non universal charge configurations naturally give rise to $Z'$ mediated flavor changing neutral currents in both the quark and lepton sectors.
We also demonstrate that, within this framework, the  $Z'$ boson can naturally acquire purely axial vector or mixed vector-axial couplings to the SM fermions, both in the heavy and light $Z'$ regimes. 

\end{abstract}

%%%%%%%%%%%%%%%%%%%%%%%%%%%%%%%%%%%%%%%%%%%%%%%%%%%%%%%%%%%

\maketitle  

\tableofcontents

%%%%%%%%%%%%%%%%%%%%%%%%%%%%%%%%%%%%%%%%%%%%%%%%%%%
\section{\label{sec:Intro}Introduction}
%%%%%%%%%%%%%%%%%%%%%%%%%%%%%%%%%%%%%%%%%%%%%%%%%%%
%%%%% 
Phenomenological studies in neutrino and B-physics often consider new light mediators or effective four-Fermi interactions, where all Lorentz-invariant bilinear currents such as scalar (S), pseudoscalar (P), vector (V), axial-vector (A), and tensor (T) are considered \cite{Sudarshan:1958vf,Feynman:1958ty,Wilson:1972ee,Kopp:2007ne,Gavela:2008ra,Alonso:2014csa,Wise:2014oea,Lindner:2016wff,Jenkins:2017jig,Silvestrini:2019sey,Babu:2019mfe,Majumdar:2021vdw,A:2022acy,Majumdar:2022nby,Majumdar:2024dms,DeRomeri:2024dbv,KATRIN:2024odq,Chattaraj:2025rtj,Chattaraj:2025fvx}.
Apart from tensor operators, the remaining operators can be realised within fully ultraviolet (UV) complete theories.
Scalar (Pseudoscalar) operators can be UV completed by extending the Standard Model with real (complex) scalar fields. In the presence of heavy mediators, such scalar Yukawa interactions can also effectively mimic vector and axial-vector operators through identities involving Fierz 
transformations  \cite{Cuypers:1996ia,Bergmann:1999pk,Nieves:2003in,Antusch:2008tz}. Consequently, pure vector, pure axial-vector, and mixed vector--axial-vector interactions have been extensively studied within the effective operator framework in the heavy mediator regime~\cite{Antusch:2008tz,Ohlsson:2012kf,AristizabalSierra:2018eqm,Amaral:2023tbs,Hurth:2023jwr,Athron:2023hmz,Hurth:2025vfx}.
%%%%
However, for light mediators one cannot use scalar Yukawa interactions to mimic vector or axial-vector interactions.
%%%%
For light mediators, vector and axial-vector interactions have to arise from new gauge symmetry.
%%%%
To realize such couplings in a model for both light and heavy mediators, 
the simplest gauge theoretic extension is to introduce an additional $U(1)_X$ gauge symmetry.
%%%%

Gauging a symmetry requires careful analysis as it may induce gauge anomalies, and their cancellation is crucial to preserving unitarity and renormalizability \cite{Adler:1969gk,Bardeen:1969md,Bell:1969ts,Delbourgo:1972xb,Witten:1982fp,Alvarez-Gaume:1983ihn}.
Gauge anomalies arise from the representations of fermions under the gauge symmetry. Vector like fermions, where left- and right-handed components share the same representation, do not contribute to gauge anomalies. An invariant Dirac mass term for such fermions is therefore allowed by the gauge symmetry~\footnote{If multiple gauge symmetries are present, such a mass term is allowed only if the fermion is vector like under all of them.}. Vector fermions also give rise to purely vector couplings with the gauge bosons associated with the corresponding symmetry.
%%%%
In contrast, chiral fermions, for which the left- and right-handed components transform differently under the gauge symmetry, can introduce gauge anomalies. Such chiral symmetries also give rise to mixed vector and axial-vector interactions and forbid gauge invariant Dirac mass terms for fermions.
%%%%
For example, the Standard Model (SM) is a chiral theory, with quarks and leptons transforming asymmetrically under $SU(2)_L \otimes U(1)_Y$. Consequently, anomaly cancellation constrains the gauge charges of fermions in the SM. The asymmetric charge assignments also lead to mixed vector and axial vector currents for the $W$ and $Z$ bosons. Moreover, a Higgs doublet charged under $SU(2)_L \otimes U(1)_Y$ is required to write gauge invariant Yukawa interactions that generate mass terms for the SM fermions.
Introducing new $U(1)_{X}$ gauge symmetries adds additional constraints on the charge assignments of SM and beyond the Standard Model (BSM) fermions. If SM fermions are chiral under $U(1)_{X}$, their mass and mixing mechanisms further constrain BSM charges. 

Coming now to the explicit $U(1)_X$ symmetries discussed in the literature, the flavor universal vector $U(1)_X$ symmetries, such as $U(1)_{B-L}$, in which the charges of all SM fermions are generation independent, preserve the SM Yukawa couplings and simplify anomaly-cancellation conditions \cite{Mohapatra:1980qe,Mohapatra:1980de,Masiero:1982fi,Mohapatra:1982xz,Buchmuller:1991ce,Basso:2008iv}. However, such minimal constructions yield purely vector couplings of the $Z'$ boson to SM fermions. 
%%%
Mixed vector and axial vector couplings can be generated by introducing heavy vector like fermions, but this generally induces tree level flavor changing neutral currents (FCNCs) in both the quark and charged-lepton sectors. In light $Z'$ scenarios, these FCNCs can be particularly problematic, especially in the charged-lepton sector.
An alternative approach employs flavor specific vector symmetries, where different SM generations carry different charges, i.e. $B-3L_i$, $B_i-3L_j$, $B_i-B_j$, $L_i-L_j$, etc., where $i,j = 1,2,3$ label fermion generations ~\cite{Ma:1997nq,He:1990pn,Appelquist:2002mw, Lee:2010hf,Ma:2014qra,Ma:2015raa,Ma:2015mjd,Bonilla:2017lsq,Alonso:2017uky,Allanach:2022blr,AtzoriCorona:2022moj, DeRomeri:2023ytt,DeRomeri:2024dbv}. 
%%%
In such cases, the mixed vector axial-vector couplings can appear in the fermion mass basis; however, this typically comes at the expense of inducing fermion mixing.
Hence, simultaneously avoiding FCNCs and generating sizable axial-vector couplings in both the charged lepton and quark sectors remains nontrivial.
%%%%%%

Another approach to generate mixed vector and axial vector couplings is to consider chiral $U(1)_X$ charge assignments for SM fermions \cite{Appelquist:2002mw,Montero:2007cd,Ma:2014qra,Oda:2015gna,Das:2016zue,Jana:2019mez,Prajapati:2024wuu}. In heavy $Z'$ scenarios, flavor universal chiral $U(1)_X$ constructions can generate mixed vector and axial-vector couplings. 
%%%%%%
However, in the light $Z'$ case the situation is more subtle.
In the light $Z'$ scenarios axial vector coupling vanishes at leading order in the gauge coupling due to the gauge invariance of the Yukawa interactions \cite{Kahn:2016vjr,DelleRose:2018eic}.  
%%%%
This suppression of the axial-vector coupling for light $Z'$ is generic for any $U(1)_X$ extension of the SM in which there is a single Higgs doublet 
whose Yukawa couplings respect $U(1)_X$ gauge invariance 
\cite{Kahn:2016vjr}.
%%%
As a consequence, minimal flavor universal chiral models, such as linear combinations of hypercharge and $B-L$, effectively reduce to purely vector like theories in the light $Z'$ regime.
%%%%
However, such suppression is not guaranteed in a two Higgs doublet models with flavor specific chiral $U(1)_X$ symmetry \cite{Kahn:2016vjr}.
%%%
Consequently, in the presence of light mediators, flavor specific chiral $U(1)_X$ constructions provide a natural and efficient framework to realize mixed vector--axial-vector or purely axial-vector couplings, as additional Higgs doublet can participate in fermion mass generation and mixing~\cite{Kahn:2016vjr,DelleRose:2017xil,Allanach:2018vjg}.
%%%%%%%%%%%%%%%%%

In this work, we propose a class of flavor specific chiral $U(1)_X$ charge assignments that satisfy all gauge anomaly cancellation conditions and naturally generate mixed vector--axial-vector or purely axial-vector couplings to SM fermions in the light $Z'$ regime.
%%%%% 
We systematically classify the anomaly free solutions and construct corresponding UV complete models under the following assumptions:
%%%%%%
\begin{itemize}
%%%%%
\item In the scalar sector, we consider two Higgs doublets, $\Phi$ and $\varphi$, together with $i$ number of SM singlet scalar fields denoted by $\chi_i$.
The singlet scalars $\chi_i$ are required for the generation of neutrino masses and mixing. Moreover, they also allow us the freedom to vary the $Z'$ mass freely.  Their exact number will be fixed in Sec.~\ref{Sec:Lep_mix} when we discuss UV complete models including neutrino mass and mixing mechanism.
\item For flavor specific scenarios, the $U(1)_X$ charges of the three generations can, in principle, be different. 
%%%%
However, due to the relatively stringent experimental limits on lepton flavor universality violation in the first two generations, arising from measurements such as $R_K$ and $R_{K^*}$ \cite{LHCb:2022qnv,LHCb:2022vje}, we assume identical $U(1)_X$ charge assignments for the first two generations of SM fermions.
%%%%
%%
\item To cancel the gauge anomalies arising from the chiral $U(1)_X$ assignments, we introduce three right-handed neutrinos. 
\item To generate neutrino masses and lepton mixing, an arbitrary number of BSM singlet scalars $\chi_{i}$ are permitted (with emphasis on minimal realizations), and no additional fermions are introduced beyond the three required for anomaly cancellation.
\end{itemize}
We present several UV complete benchmark models illustrating the range of possible charge assignments, under which the quark and lepton flavor structures can differ substantially, leading to distinct phenomenological signatures. 
%%%%
We demonstrate the viability of these models in obtaining 
phenomenologically interesting purely axial-vector and mixed 
vector axial-vector couplings within the light $Z'$ regime. In this setup, we show how neutrino scattering bounds can be effectively evaded. Moreover, in the heavy $Z'$ regime, we demonstrate that such models can address the $B$-physics anomalies observed in both angular observables and branching fractions for decays into charged leptons.

The paper is organized as follows. In Sec.~\ref {Sec:Mass_spectrum_of_gauge_mediators}, we discuss the gauge boson mass spectrum, the phenomenology of the $Z$-$Z'$ mixing angle, and constraints from the $\rho$ parameter in both light and heavy $Z'$ scenarios. 
Sec.~\ref{Sec:Interactions} examines the interactions of the gauge bosons with fermions, including the structure of vector and axial-vector couplings in both the heavy- and light-$Z'$ regimes.
In Sec.~\ref{Sec:anomaly_cancellations}, we present the solutions to the gauge anomaly cancellation conditions and classify them according to the mechanisms for SM fermion mass generation.
Sec.~\ref{Sec:Quark_mixing} addresses the generation of quark mixing in models with one or two Higgs doublets. Finally, Sec.~\ref{Sec:Lep_mix} discusses lepton mixing and possible UV completions of the Weinberg operator to generate Majorana neutrino masses.
The phenomenological implications of these models, including the conditions under which pure axial-vector couplings of the $Z'$ to SM fermions arise and their relevance for B-physics anomalies and neutrino interactions, are examined in Sec. \ref{Sec:Phenomenological Implications}. Finally, in Sec. \ref{Sec:Conclusion}, we provide concluding remarks.
%%%%%%%%%%%%%%%%%%
%%%%%%%%%%%%%%%%%%
%%%%%%%%%%%%%%%%%%
%%%%%%%%%%%%%%%%%%
\section{Mass Spectrum of Gauge Mediators} \label{Sec:Mass_spectrum_of_gauge_mediators}
%%%%%%%%%%%%%%%%%%
%%%%%%%%%%%%%%%%%%
%%%%%%%%%%%%%%%%%%
%%%%%%%%%%%%%%%%%%
In this section, we discuss the masses of the gauge bosons after spontaneous symmetry breaking (SSB).
To keep this discussion as model independent as possible, we fix only the scalar sector here. Consequently, the analysis remains independent of any other specific model details.
Since the $U(1)_X$ charges of the scalars are not constrained by anomaly cancellation conditions, in order to maintain generality, we treat them as free parameters throughout this section.
Constraints on their $U(1)_X$ charge assignments arising from fermions mass generation mechanism are discussed in Sec.~\ref{Sec:anomaly_cancellations}.

As outlined earlier, the scalar content consists of two Higgs doublets, $\Phi$ and $\varphi$, together with $i$ number of SM singlet scalar fields denoted by $\chi_i$. The singlet scalars $\chi_i$ are required to generated masses and mixing of the neutrinos. In addition, they also allow the $Z'$ mass to vary freely.
%%%%%%%
In the later section on neutrino masses (see Sec.~\ref{Sec:Lep_mix}), we will fix their number as well, but for the time being we keep them as free parameters.
%%%%%%%
The mass spectrum of gauge bosons arises from the expansion of the kinetic terms of the scalar fields after SSB of both the electroweak and $ U(1)_{X} $ symmetries. The kinetic terms of scalars are written as,
%%%%%
\begin{equation}\label{Eq:Kinetic_term_scalar}
(D_{\mu}\Phi)^{\dagger}D^{\mu}\Phi+ (D_{\mu}\varphi)^{\dagger}D^{\mu}\varphi+(D_{\mu}\chi_{i})^{\dagger}D^{\mu}\chi_{i}~.
\end{equation}
%%%%%%
The covariant derivative, $D_{\mu}$, in this case is defined as,
%%%%%
\begin{equation}\label{codrivative}
D_{\mu}= \partial_{\mu} +igT^{a}W^{a}_{\mu} + ig'\frac{Y}{2}B_{\mu}+ig_xXC_{\mu}\,.
\end{equation}
%%%%%
In this context, $g$, $g'$, and $g_x$ denote the gauge couplings corresponding to  $SU(2)_{L}$, $U(1)_{Y}$ and $U(1)_{X}$, respectively, while the charges associated to $U(1)_{Y}$ and $U(1)_{X}$ are represented by $Y$ and $X$, respectively. The generators of $SU(2)_{L}$ are given by $T^{a} = \sigma^{a}/2$, where $\sigma^{a}$ represents the Pauli matrices.
%%%%%%%%%
The electrically charged gauge bosons are defined as $W_{\mu}^{\pm} = (W_{\mu}^{1} \mp i\, W_{\mu}^{2})/\sqrt{2}$, and the corresponding generators are $T^{\pm} = (T^{1} \pm T^{2})/\sqrt{2}$.
%%%%%%%%%
%%%%%%%%%
Both the Higgs doublets, $\Phi$ and $\varphi$, and the SM singlet scalars, $\chi_i$, acquire vacuum expectation values (VEV), thereby breaking the electroweak and $U(1)_X$ symmetries. The corresponding VEV are given by,
%%%%%%%%%%
\begin{equation}\label{VEV}
\langle \Phi \rangle = \frac{1}{\sqrt{2}}\begin{bmatrix}
0 \\
v_{\Phi}
\end{bmatrix},~~~~\langle \varphi \rangle = \frac{1}{\sqrt{2}}\begin{bmatrix}
0 \\
v_{\varphi}
\end{bmatrix},~~~~ \langle \chi_{i} \rangle = \frac{v_{i}}{\sqrt{2}}~.
\end{equation} 
%%%%%%%%%%%%%%%%%%%%%%%%%%%%%%%%%%%%%%%%%%%%%%%%%%%%%%%%%%%%%%%%%%%%%%%%%%%%%%%%%%%
By substituting the covariant derivative and fields with the expression defined in Eq.~\eqref{codrivative} and Eq.~\eqref{VEV}, we obtain the gauge boson mass matrices.
%%%%%%%%
Since there is no mixing in the charged sector, the $W$ boson mass remains identical to its SM expression, $M_{W}^{2} = g^{2} v^{2}/4$, with $v = \sqrt{v_{\Phi}^{2} + v_{\varphi}^{2}} \approx 246.22~\text{GeV}$. The neutral gauge bosons, however, mix with one another, and the corresponding mass matrix in the basis $(B_{\mu}, W_{\mu}^{3}, C_{\mu})$ takes the form,
%%%%%%%%
\begin{equation}\label{Eq:Gauge_Boson_Mass_mat}
\mathcal{M}^2_{_{V}}= \frac{v^{2}}{4}\begin{pmatrix}
g'^{2} & -gg' & 2qg'g_x\\
-gg'   &  g^{2} & -2qgg_x\\
2qg'g_x & -2qgg_x & 4u^{2}g_x^{2}
\end{pmatrix}\,.
\end{equation}
%%%%%%%%%%%%%%%%%%%%%%%%%%%%%%%
Where,
\begin{eqnarray}\label{Eq:Exact_Mass_mat_para}
&& v^{2} = v_{\Phi}^{2} + v_{\varphi}^{2} \approx (246.22~ \text{GeV})^2\,,~~~~ q = X_{\Phi} + \frac{v_{\varphi}^{2}}{v^{2}} (X_{\varphi} - X_{\Phi}) \,, \\  && u^{2}=X_{\Phi}^{2}+\frac{u_{\chi}^{2}}{v^{2}} + \frac{v_{\varphi}^{2}}{v^{2}} (X_{\varphi}^{2} - X_{\Phi}^{2})\,.
\end{eqnarray}  
%%%%%%%%%%%%%%%%%%%%%%%%
with $u_{\chi}$  defined as $u_{\chi}=\sqrt{\sum_{i}(X^{^{2}}_{\chi_{_{i}}}v_{i}^{2})}$. Here, $X_{\chi_i}$ denotes the $U(1)_X$ charge of the singlet scalar $\chi_i$. Similarly, $X_\Phi$ and $X_\varphi$ denote the $U(1)_X$ charges of the Higgs doublets $\Phi$ and $\varphi$, respectively.
%%%%%
Notice that, in the limit $v_{\varphi}^{2} \ll v_{\Phi}^{2}$, these parameters can be approximated as follows,
\begin{equation}
    q \simeq X_{\Phi},~~~ u^{2} \simeq X_{\Phi}^{2}+\frac{u_{\chi}^{2}}{v^{2}}\,.
\end{equation}
In this limit, the gauge boson mass matrix of Eq.~\eqref{Eq:Gauge_Boson_Mass_mat} reduces to the single Higgs doublet scenario. 
This limit ensures that the lightest CP-even scalar has SM-like couplings to the gauge bosons.
For the remainder of this work, we restrict our analysis to this limit, corresponding effectively to the alignment regime \cite{Branco:2011iw}.

We proceed by diagonalizing the mass matrix given in Eq.~\eqref{Eq:Gauge_Boson_Mass_mat} with the aid of an orthogonal matrix $\mathcal{O}(\alpha)$.  
The gauge eigenstates $(B^{\mu}, W_{3}^{\mu}, C^{\mu})$ and the mass eigenstates $(A^\mu, Z^\mu, Z^{\prime \mu})$ are related by,
%%%%%%%%%%%%
\begin{equation}
\label{unitary matrix}
\begin{bmatrix}
A^{\mu} \\
Z^{\mu} \\
Z^{\prime \mu}
\end{bmatrix} = \mathcal{O}(\alpha) \begin{bmatrix}
B^{\mu} \\
W_{3}^{\mu}\\
C^{\mu}
\end{bmatrix}.
\end{equation}
%%%%%%  
The resulting diagonal mass matrix is
\begin{equation}\label{Eq:Gauge_and_mass_eigen_states}
\mathcal{M}^2_{\text{diag}} = \mathcal{O}(\alpha)\, \mathcal{M}^2_{V}\, \mathcal{O}(\alpha)^{\dagger}\,,
\end{equation}
%%%%%%
and the rotation matrix is defined as,
\begin{equation}\label{Eq:O_alpha_Matrix}
    \mathcal{O}(\alpha) = \begin{bmatrix}
\cos\theta_{W} &~ \sin\theta_{W} &~0\\
-\cos\alpha \sin\theta_{W} & \cos\alpha \cos\theta_{W}
&~ -\sin \alpha\\
-\sin\alpha \sin\theta_{W} &~  \sin\alpha\cos\theta_{W} &~ \cos\alpha 
\end{bmatrix}\,.
\end{equation}
Here, $\theta_{W}$ denotes the weak mixing angle, while $\alpha$ governs the mixing between the $Z$ and $Z'$ bosons. 
%%%%%%%%%%
The mixing between the $Z'$ and the photon is assumed to be negligible and is therefore set to zero.
%%%%%%%%%
Following the rotation to mass basis, one mass eigenstate becomes zero which is identified as the photon, while the remaining two mass eigenstates read
%%%%%%%%
\begin{equation}\label{Eq:Gauge_boson_mass}
\begin{aligned}
M_{\text{Light}}^{2}= \frac{v^{2}}{8}(A_{0}-\sqrt{B_{0}^{2}+C_{0}^{2}}),&&&&&&\,M_{\text{Heavy}}^{2}=\frac{v^{2}}{8}(A_{0}+\sqrt{B_{0}^{2}+C_{0}^{2}})\,,
\end{aligned}
\end{equation}
%%%%%%%%%%%%%%%%
with $A_0$, $B_0$ and $C_0$ defined as,  
%%%%%%%%%%%%%%%%
\begin{equation}\label{Eq:A0_B0_C0}
\begin{aligned}
      A_{0}= g^{2}+{g'}^{2}+4u^{2}g_x^{2} \,,~ ~   B_{0}= 4X_{\Phi}g_x\sqrt{g^{2}+{g'}^{2}} \,,~ ~
      C_{0}=  g^{2} + {g'}^{2} - 4u^{2}g_x^{2}\,.
      \end{aligned}
\end{equation}
%%%%%%%%%%%%%%%%
The $u_{\chi}$ parameter can be expressed as a function of $M_{Z'}$ and $g_x$,
%%%%%%
\begin{equation}
\label{Eq:VEV_Paramter_ux}
u_{\chi} = \frac{M_{Z'}}{g_x}\sqrt{   \frac{v^{2} (g^{2} + g'^{2} + 4 g_x^{2}X_{\Phi}^{2}) - 4M_{Z'}^{2}}{v^{2} (g^{2} + g'^{2} ) - 4M_{Z'}^{2}} } =  \frac{M_{Z'}}{g_x}\sqrt{   \frac{ [(M_{Z}^{\text{SM}})^{2} + v^{2}g_x^{2}X_{\Phi}^{2}] - M_{Z'}^{2}}{(M_{Z}^{\text{SM}})^{2} - M_{Z'}^{2} }  }\,. 
\end{equation}
%%%%
Here, $M_{Z}^{\text{SM}}$ is the mass of $Z$
boson in SM at tree level, defined as $M_{Z}^{\text{SM}} = v\sqrt{g^{2}+g'^{2}}/2$.
%%%%%%
From Eq.~\eqref{Eq:Gauge_boson_mass}, it is evident that in the limit $u_\chi \to 0$, the lighter eigenstate $M_{\text{Light}}$ is of order $\mathcal{O}(v_\varphi g_x)$. Therefore, the presence of at least one scalar singlet is required to allow the mass of $Z'$ to vary independently of the electroweak scale.
%%%%%%

The mixing among the neutral gauge bosons leads to two possible mass hierarchies: $M_{Z'} < M_{Z}$ and $M_{Z'} > M_{Z}$ between $Z$ and $Z'$. The parametrization of $\alpha$ changes sign when changing the mass hierarchy i.e. if one takes $+\alpha$ when $M_{Z'} < M_{Z}$, then for $M_{Z'} > M_{Z}$ one should take $\alpha \to -\alpha$ in Eq. \eqref{Eq:O_alpha_Matrix}. Each hierarchy corresponds to a distinct physical regime, which we examine in the following subsections.
%%%%%%%%%%%%%%%%%%%%%%%%%%%%%%%%%%%%%%%%%%%%%%%
%%%%%%%%%%%%%%%%%%%%%%%%%%%%%%%%%%%%%%%%%%%%%%%

\subsection{Light $Z'$ scenario: $\mathbf{M_{Z'} < M_{Z}}$} Here we discus $Z'$ mass, mixing angle and their phenomenology in the limit $M_{Z'} < M_{Z}$, i.e., $M_{Z'(Z)} \equiv M_{\text{Light(Heavy)}}$. 
In this case, the diagonalization of the mass matrix in Eq.~\eqref{Eq:Gauge_Boson_Mass_mat} is performed by the orthogonal matrix $\mathcal{O}(\alpha)$ defined in Eq.~\eqref{Eq:O_alpha_Matrix}. 
The rotation angles are defined as,
%%%%
\begin{equation}
\label{Eq:angle}
\tan\theta_{W} = \frac{g'}{g},~~~ \sin 2\alpha = \frac{4X_{\Phi}g_x}{ \sqrt{g^{2}+{g'}^{2}} } \frac{(M_{Z}^{\text{SM}})^{2}}{M_{Z}^{2} - M_{Z'}^{2}} \,.
\end{equation}
%%%%%%
The mixing between the neutral gauge bosons induces a shift in the $Z$ boson mass proportional to the mixing angle. In the limit $\alpha \to 0$, this correction vanishes, and the $Z$ boson mass reduces to its SM value, $M_{Z} \to M_{Z}^{\text{SM}}$.
%%%%
The mixing angle $\alpha$ could be constrained by the precisely measured electroweak parameter $\rho = M_{W}^{2}/(M_{Z}^{2}\cos^{2}\theta_{W})$ \cite{Ross:1975fq,Bento:2023weq}. This parameter is 1 at tree level in SM. In $SM \otimes U(1)_X$ theories, however, it deviates from unity and can be expressed as~\cite{Bento:2023flt},
\begin{equation}
\label{Eq:Rho_parameter}
    \rho -1 = \left[  \left( \frac{M_{Z'}}{M_{Z}} \right)^{2} -1    \right] \sin^{2} \alpha\,.
\end{equation}
Using the $3\sigma$ allowed range of the $\rho$ parameter, $\rho = 1.00038 \pm 0.00060$ \cite{ParticleDataGroup:2020ssz}, obtained from the most recent global fit to electroweak precision data, gives constraint $X_{\Phi} g_x \lesssim 5.5 \times 10^{-3}$ for sufficiently light $Z'$ ($M_{Z'}^{2}/M_{Z}^{2} << 1$) \cite{Majumdar:2024dms}.
%%%%%%
Under this limit, Eqs.~\eqref{Eq:VEV_Paramter_ux}, \eqref{Eq:angle}, and \eqref{Eq:Rho_parameter} can be approximated as,
\begin{equation}\label{Eq:approx_light_zp}
    u_{\chi} \simeq \frac{M_{Z'}}{g_x}\,,~~~ \alpha \simeq \frac{2 \cos{\theta_{W}}  }{g} X_{\Phi}g_x\,,~~~ \rho \simeq 1 - \alpha^2\,.
 \end{equation}
Interestingly, in this limit, the mixing angle $\alpha$ becomes independent of $Z'$ mass. 
\begin{figure}[ht]
\begin{center}
\includegraphics[width=0.40\linewidth]{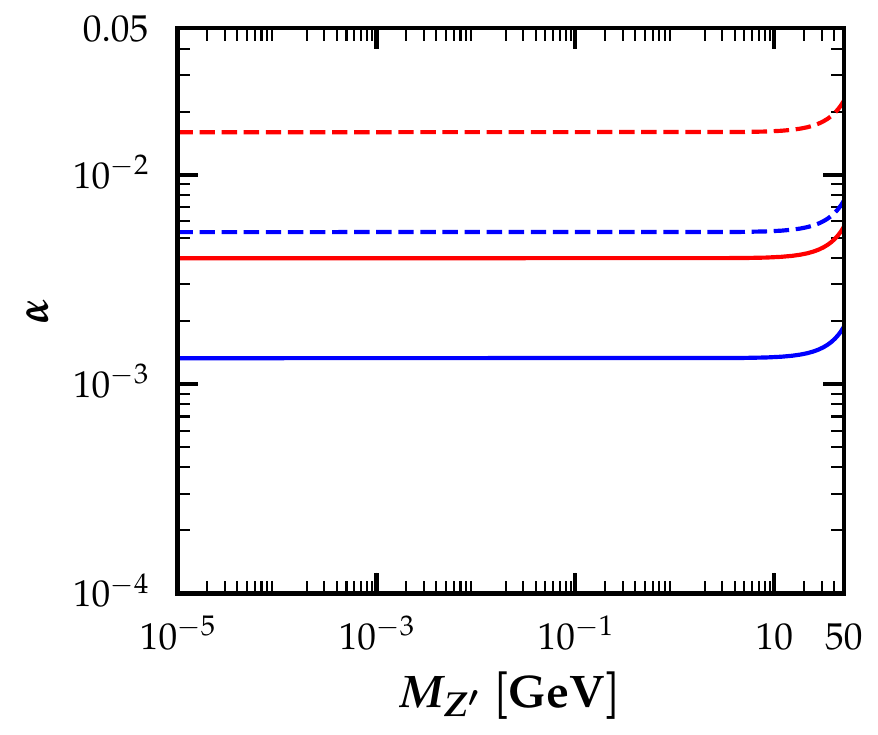}
\includegraphics[width=0.58\linewidth]{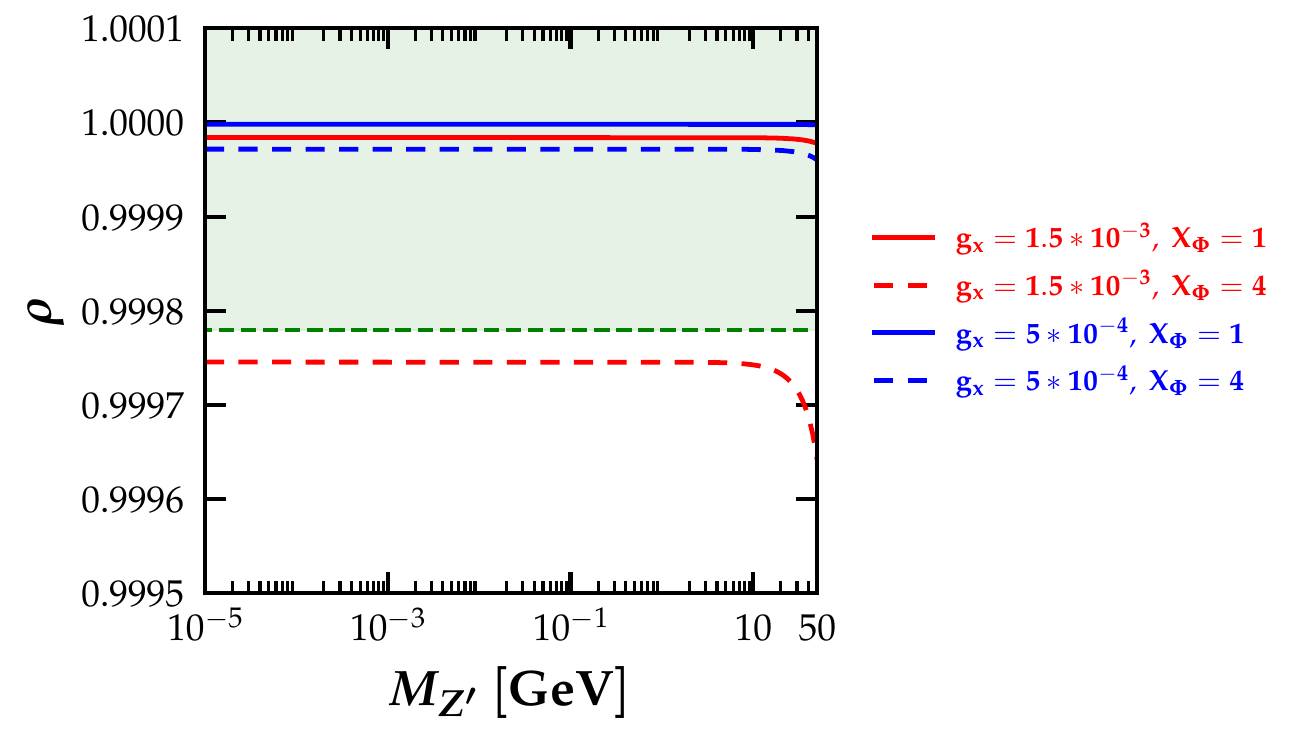}
\end{center}
\caption{Left (Right)  panel shows mixing angle ($\rho$ parameter) as a function of $M_{Z'}$ for fixed values of gauge coupling $g_x$ and Higgs charge $X_{\Phi}$. The green band shows experimentally measured value of $\rho$ parameter at $3\sigma$, $\rho = 1.00038 \pm 0.00060$ \cite{ParticleDataGroup:2020ssz}.}
\label{FiG:Alpha_Light_Zp}
\end{figure}
%%%%%%%%%
%%%%%%%%%
Fig.~\ref{FiG:Alpha_Light_Zp} left panel shows the mixing angle $\alpha$ as a function of the $Z'$ mass $M_{Z'}$ for different values of the gauge coupling $g_x$, plotted using expression \eqref{Eq:angle}. 
%%%
The solid (dashed) red and blue lines correspond to two different values of the gauge coupling for $X_{\Phi}=1~(4)$.
%%%
The right panel of Fig.~\ref{FiG:Alpha_rho_Heavy_Zp}, we show the $\rho$ parameter as a function of $M_{Z'}$ for the same benchmark values of $X_{\Phi}$ and $g_{x}$.
%%%
It is evident from the figure that even for $M_{Z'}$ as large as $10~\mathrm{GeV}$, both $\alpha$ and $\rho$ remains independent of the $Z'$ mass.
%%%
Thus, the approximate expressions 
given in Eq.~\eqref{Eq:approx_light_zp} provide a very good  approximation for light $Z'$ bosons with $M_{Z'} \lesssim 10~\mathrm{GeV}$.
%%%

%%%%%%%%%%%%%%%%%%%%%%%%%%%%%%%%%%%%%%%%%%%%%%%%%
%%%%%%%%%%%%%%%%%%%%%%%%%%%%%%%%%%%%%%%%%%%%%%%%%
\subsection{Heavy $Z'$ scenario: $\mathbf{M_{Z'} > M_{Z}}$ }\label{Sub:sec:Heavy $Z'$ : mass and couplings}
\begin{figure}[ht]
\begin{center}
\includegraphics[width=0.40\linewidth]{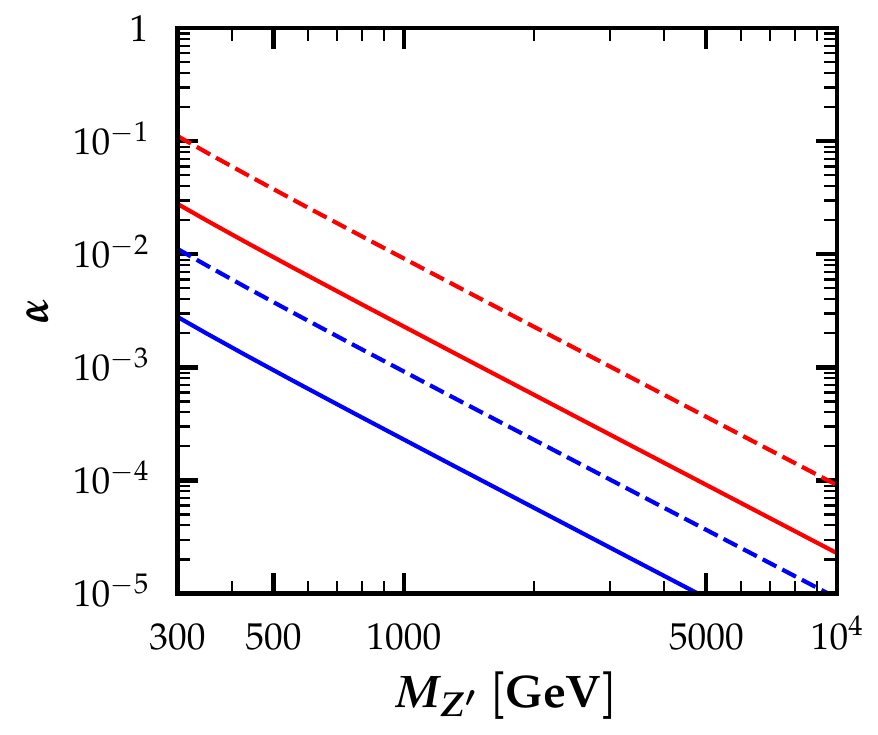}
\includegraphics[width=0.58\linewidth]{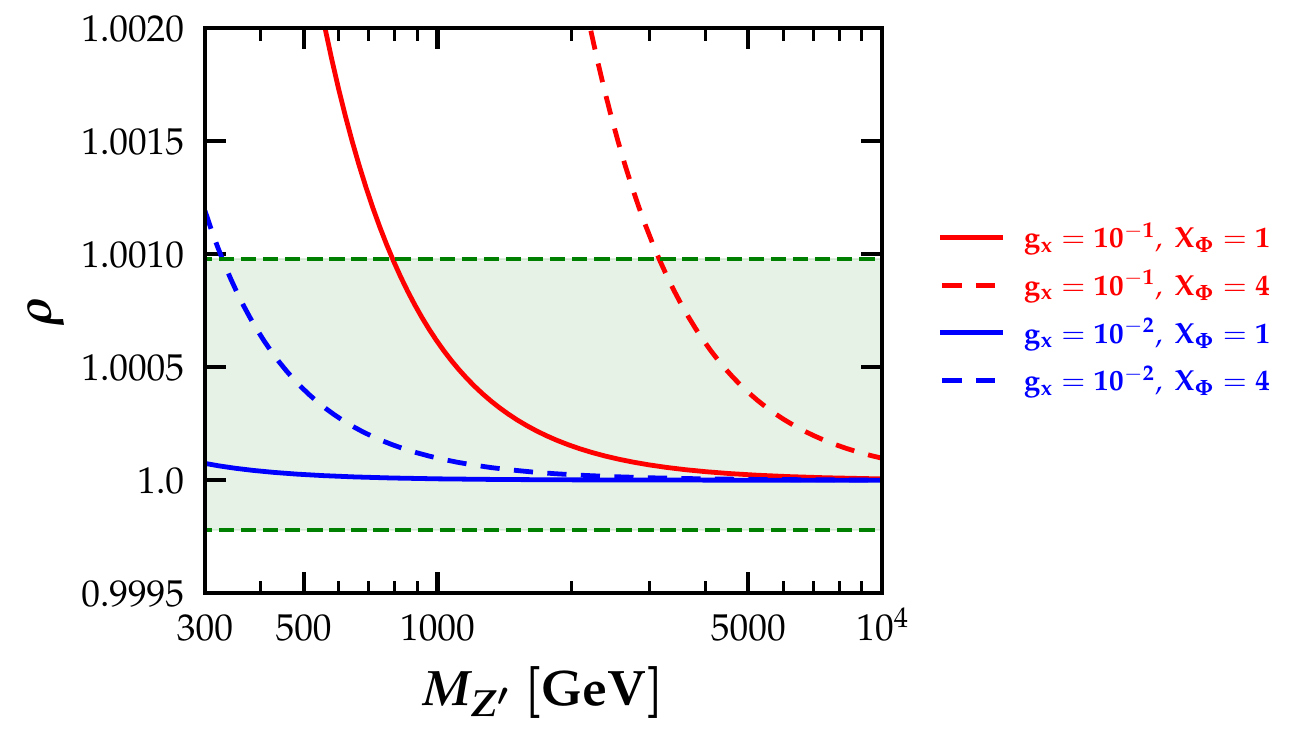}
\end{center}
\caption{Left (Right)  panel shows mixing angle ($\rho$ parameter) as a function of $M_{Z'}$ for fixed values of gauge coupling $g_x$ and Higgs charge $X_{\Phi}$. The green band shows experimentally measured value of $\rho$ parameter at $3\sigma$, $\rho = 1.00038 \pm 0.00060$ \cite{ParticleDataGroup:2020ssz}.}
\label{FiG:Alpha_rho_Heavy_Zp}
\end{figure}

%%%%%%%%%%%%%%%%%%%%%%%%%%%%%%%%%%%%%%%%%%%%%%%%%
%%%%%%%%%%%%%%%%%%%%%%%%%%%%%%%%%%%%%%%%%%%%%%%%%
Here we discuss the $Z'$ mass, mixing angle and their properties in the limit $M_{Z'} > M_{Z}$, i.e., $M_{Z(Z')} \equiv M_{\text{Light(Heavy)}}$. In this case, the diagonalization of the mass matrix in Eq.~\eqref{Eq:Gauge_Boson_Mass_mat} is performed using the orthogonal matrix $\mathcal{O}(-\alpha)$, where $\alpha$ is replaced by $-\alpha$ in Eq.~\eqref{Eq:O_alpha_Matrix}. The gauge and mass eigenstates are related as shown in Eq.~\eqref{Eq:Gauge_and_mass_eigen_states}. After diagonalization, the masses of the gauge bosons are given by Eq. \eqref{Eq:Gauge_boson_mass} with $M_{\text{Light}}=M_{Z}$ and $M_{\text{Heavy}}=M_{Z'}$.
%%%
The expressions for $A_{0}$, $B_{0}$, and $C_{0}$ remain same as Eq.~\eqref{Eq:A0_B0_C0}, while the expression for $u_{\chi}$ remains same as in Eq.~\eqref{Eq:VEV_Paramter_ux}. The expression for $\sin 2\alpha$ can be obtained from Eq.~\eqref{Eq:angle} by making the replacement $\alpha \to -\alpha$. The expression for the $\rho$ parameter also remains the same as in Eq.~\eqref{Eq:Rho_parameter}. In heavy $Z'$ scenario, no simple approximate expressions analogous to Eq.~\eqref{Eq:approx_light_zp} are available for $\rho$ and $\alpha$, and both remain functions of $g_x$, $M_{Z'}$, and $X_{\Phi}$, as shown in the Fig.~\ref{FiG:Alpha_rho_Heavy_Zp}.
In the left panel of Fig.~\ref{FiG:Alpha_rho_Heavy_Zp}, we have plotted the mixing angle $\alpha$ as a function of the $Z'$ mass $M_{Z'}$ for fixed values of the gauge coupling $g_{x}$ and the Higgs charge $X_{\Phi}$. The solid (dashed) red and blue lines correspond to two different values of the gauge coupling for $X_{\Phi}=1~(4)$. It is clear from the figure that for fixed values of $g_{x}$ and $X_{\Phi}$, increasing the $Z'$ mass leads to a decrease in the mixing angle $\alpha$, which approaches zero in the limit $M_{Z'} \to \infty$. Furthermore, increasing the Higgs charge $X_{\Phi}$ and the gauge coupling $g_{x}$ enhances the mixing.
In the right panel of Fig.~\ref{FiG:Alpha_rho_Heavy_Zp}, we show the $\rho$ parameter as a function of $M_{Z'}$ for the same benchmark values of $X_{\Phi}$ and $g_{x}$. 
The green band shows experimentally measured value of $\rho$ parameter at $3\sigma$, $\rho = 1.00038 \pm 0.00060$ \cite{ParticleDataGroup:2020ssz}. 
Since the $Z$ mass is smaller than the SM predicted value in this scenario, the $\rho$ parameter becomes larger than unity. Consequently, high values of $g_{x}$ and $X_{\Phi}$ at low $Z'$ masses are excluded by constraints on the $\rho$ parameter, as shown in Fig. \ref{FiG:Alpha_rho_Heavy_Zp}.

Having discussed the gauge boson mass spectrum and the properties of the $Z$--$Z'$ mixing angle $\alpha$, we now turn to the interactions of these bosons with SM fermions, including the role of $\alpha$ in modifying these couplings.

%%%%%%%%%%%%%%%%%%%%%%%%%%%%%%%%%%%%%%%%%%%%%%%%%%%%%%%%%%%%%%%%%%%%%%%%%%%%%%%%%%%%%%
\section{Interactions between fermions and gauge bosons}\label{Sec:Interactions}
%%%%%%%%%%%%%%%%%%%%%%%%%%%%%%%%%%%%%%%%%%%%%%%%
%%%%%%%%%%%%%%%%%%%%%%%%%%%%%%%%%%%%%%%%%%%%%%%%
In this section, we discuss the interactions of the gauge bosons with the SM fermions.
Again, to keep the discussion as general as possible, the fermion 
charges are treated as free parameters throughout this section.
Constraints on fermion charge assignments arising from the gauge anomaly cancellation are discussed in Sec.~\ref{Sec:anomaly_cancellations}.
The interactions between gauge bosons and fermions arise from the
covariant derivative defined in Eq.~\eqref{codrivative}. However,
mixing among the gauge bosons alters the couplings of the physical gauge bosons to fermions.
As discussed in Sec.~\ref{Sec:Mass_spectrum_of_gauge_mediators}, the choice of the rotation matrix in Eq.~\eqref{unitary matrix} ensures that the photon does not mix with the new gauge boson $Z'$. Consequently, QED interactions remain identical to those in the SM. Furthermore, since no additional mixing occurs in the charged sector, the $W$ boson mass and the charged current weak interactions remains the same as SM. However, in the neutral current sector the interactions will deviate from SM expression due to mixing.
%%%

Following the rotation of the gauge bosons to the mass basis defined in Eq.~\eqref{unitary matrix}, the full covariant derivative can be expressed as:
%%%%%%%%%%%%%%%%%%%%%%%%%%%%%%%%%%%%%%%
\begin{equation}\label{Eq:codrivative2}
\begin{split}
D_{\mu}= &~ \partial_{\mu} +D_{\mu}^{\text{WCC}} + D_{\mu}^{\text{QED}}+ D_{\mu}^{\text{WNC}} +  D_{\mu}^{\text{BSM}}\,,
\end{split}
\end{equation}
where,
%%%%%%%%%%%%%%%%%%%%%%%%%%%%%%%%%%%%%%%
\begin{subequations}
\label{Eq:codrivative3}
\begin{align}
D_{\mu}^{\text{WCC}} &~= ig \left[ T^{+} W_{\mu}^{+} + T^{-}W_{\mu}^{-} \right],\,~~~~ D_{\mu}^{\text{QED}} = ieQA_{\mu}\,,  \\
D_{\mu}^{\text{WNC}} &~= i\frac{g}{\cos \theta_{W}} \left[ (T^{3} - Q\sin^{2} \theta_{W}) \cos \alpha - X\frac{\cos \theta_{W}}{g} g_{x}\sin \alpha   \right] Z_{\mu}\,, \\
 D_{\mu}^{\text{BSM}} &~= i \left[  \frac{g}{\cos \theta_{W}} (T^{3} - Q \sin^{2} \theta_{W}) \sin \alpha + X g_{x} \cos \alpha \right] Z'_{\mu}\,.
\end{align}
\end{subequations}
%%%%%%%%%%%%%%%%%%%%%%%%%%%%%%%%%%%%%%%
%%%%%%%%%%%%%%%%%%%%%%%%%%%%%%%%%%%%
Here $T^3$ and $Q$ represent the third component of the weak isospin and the electric charge of a fermion, and $X$ represents its  $U(1)_X$ charge. 
From Eq. \eqref{Eq:codrivative3} it is evident that the QED and charged current weak interactions remains the same as SM. The Lagrangian density corresponding to these interaction in the gauge basis of fermions is given as,
%%
%%%%%%%%%%%%%
\begin{equation}
    -\mathscr{L}_{\text{WCC,QED}} = \frac{g}{\sqrt{2}} \left(  \overline{e_{_\mathtt{L}}} \gamma^{\mu} \nu_{_\mathtt{L}}  + \overline{d_{_\mathtt{L}}}\gamma^{\mu}u_{_\mathtt{L}} \right)W_{\mu}^{-}  + eQ\, \Bar{\psi}\gamma^{\mu} \psi A_{\mu} +  h.c.\,.
\end{equation}
%%%%%%%%%%%%%
In the neutral sector, however, gauge boson mixing modifies the weak neutral currents.
%%%%%%%%%%%%%
The Lagrangian density corresponding to the neutral current interactions mediated by the $Z$ and $Z'$ bosons is given by,
%%%%%%%%%%%%%
\begin{equation} \label{SM_lagrangian}
- \mathscr{L}_{\text{NC}} =  \frac{g}{\cos\theta_{W}} \overline{\psi} \gamma^{\rho}\left(g_{\psi_{_{\mathtt{L}}}}^{z} P_{\mathtt{L}} + g_{\psi_{_{\mathtt{R}}}}^{z} P_{\mathtt{R}}\right)\psi Z_{\mu}+ \overline{\psi} \gamma^{\mu}\left(g_{\psi_{_{\mathtt{L}}}}^{z'} P_{\mathtt{L}}+g_{\psi_{_{\mathtt{R}}}}^{z'} P_{\mathtt{R}}\right)\psi ~Z'_{\mu} \,. 
\end{equation} 
%%%%%%%%%%%%%%%%%%%%
The couplings $g_{\psi_{_{\mathtt{L/R}}}}^{z}$ and $g_{\psi_{_{\mathtt{L/R}}}}^{z'}$ in the fermion gauge basis for $M_{Z'} < M_{Z}$  are given by,
%%%%%%%%%%%%%%%%%%%
\begin{subequations}\label{SM Couplings}
\begin{align}
&g_{\psi_{_{\mathtt{L}}}}^{z} = \Big(T_{\psi_{_{\mathtt{L}}}}^3 - Q_{\psi}\sin^{2}\theta_{W}\Big)\cos{\alpha} - \frac{X_{\psi_{_{\mathtt{L}}}}g_x}{g}\sin{\alpha}\cos{\theta_{W}}\,,\label{SM Couplings1} \\
&g_{\psi_{_{\mathtt{R}}}}^{z} = - Q_{\psi}\sin^{2}\theta_{W}\cos{\alpha} - \frac{X_{\psi_{_{\mathtt{R}}}}g_x}{g}\sin{\alpha}\cos{\theta_{W}}\, ,\label{SM Couplings2}\\
& g_{\psi_{_{\mathtt{L}}}}^{z'} =  \frac{g}{\cos \theta_{W}} \Big(T_{\psi_{_{\mathtt{L}}}}^3 - Q_{\psi} \sin^{2} \theta_{W} \Big)\sin{\alpha} + X_{\psi_{_{\mathtt{L}}}}g_x\cos{\alpha}\,,\label{SM Couplings3} \\
& g_{\psi_{_{\mathtt{R}}}}^{z'} =   -\frac{g}{\cos \theta_{W}} \Big( Q_{\psi} \sin^{2} \theta_{W} \Big)\sin{\alpha} + X_{\psi_{_{\mathtt{R}}}}g_x\cos{\alpha}\,\label{SM Couplings4}.
\end{align}
\end{subequations}
%%%%%%%%%%%%%%%%
Where $T_{\psi_{_{\mathtt{L}}}}^3$ and $Q_{\psi}$ represent the third component of the weak isospin and the electric charge of the fermion  $\psi$. 
%
%%%%%%%%%%%
Furthermore, for the case of $M_{Z'} > M_{Z}$, one should substitute $\alpha \to -\alpha$ in Eqs. \eqref{SM Couplings1} - \eqref{SM Couplings4}. 
%%%
Notice that Z boson couplings to fermions have both SM and BSM pieces due to mixing between bosons. The additional contribution is proportional to the gauge coupling $g_x$ and the rotation angle $\alpha$.
%%%
Using Eq.~\eqref{SM_lagrangian},
the vector and axial vector couplings of the fermions with the $Z'$ is defined as, 
\begin{equation}\label{Eq:Vector_axialVector_Coupling}
    C^{\psi}_{V} = \frac{g_{\psi_{_{\mathtt{L}}}}^{z'} + g_{\psi_{_{\mathtt{R}}}}^{z'}}{2},~~~~ C^{\psi}_{A} =  \frac{g_{\psi_{_{\mathtt{L}}}}^{z'} - g_{\psi_{_{\mathtt{R}}}}^{z'}}{2}\,.
\end{equation}
%%%%%%
The couplings $g_{\psi_{_{\mathtt{L}}}}^{z'}$ and $g_{\psi_{_{\mathtt{R}}}}^{z'}$ in fermionic gauge basis for case $M_{Z'} < M_{Z}$ are defined in Eq. \eqref{SM Couplings}. The corresponding vector and axial vector coupling is given by,
%%%%%%%%%%%%%%%%%%%
%%%%%%%%%%%%%%%%%%%
\begin{subequations}\label{Eq:AVCouplings_Zp}
\begin{align}
& C^{\psi}_{V} =  \frac{g}{2\cos \theta_{W}} \Big(T_{\psi_{_{\mathtt{L}}}}^3 - 2Q_{\psi} \sin^{2} \theta_{W} \Big)\sin{\alpha} + C^{\psi}_{V(0)}\,g_x \cos{\alpha} \,, \\
%%%%%
& C^{\psi}_{A} =  \frac{g}{2\cos \theta_{W}} T_{\psi_{_{\mathtt{L}}}}^3 \sin{\alpha} + C^{\psi}_{A(0)}\,g_x \cos{\alpha}\,.
\end{align}
\end{subequations}
%%%%%%%%%%%%%%%%
Here, $C^{\psi}_{V(0)/A(0)} = (X_{\psi_{_{\mathtt{L}}}} \pm X_{\psi_{_{\mathtt{R}}}})/2$ denote the vector and axial-vector charges in the limit of vanishing mixing angle $\alpha \to 0$. As before for $M_{Z'} > M_{Z}$ one should substitute $\alpha \to - \alpha$ in Eq. \eqref{Eq:AVCouplings_Zp}.
%%%%%%%
%%%%
Since $\alpha$ exhibits 
different phenomenological behavior in the light and heavy $Z'$ regimes, we discuss the phenomenology of these couplings separately 
for the light and heavy $Z'$ cases in the following subsections.
%%%%%%%

%%%%%%%
%%%%%%%%%%%%%%%%%%%%%%%%%%%%%%%%%%
%%%%%%%%%%%%%%%%%%%%%%%%%%%%%%%%%%
\subsection{Light $Z'$ scenario: $\mathbf{M_{Z'} < M_{Z}}$}  
%%%%%%%%%%%%%%%%%%%%%%%%%%%%%%%%%
%%%%%%%%%%%%%%%%%%%%%%%%%%%%%%%%%
In the case of a light $Z'$, the couplings of SM fermions to the $Z$ and $Z'$ bosons are given in Eq.~\eqref{SM Couplings}.
We begin with the Z boson couplings. As seen from 
Eqs.~\eqref{SM Couplings1} and \eqref{SM Couplings2}, the $Z$ boson 
coupling receives both SM and BSM contributions. 
%%%
The BSM contribution is governed by the gauge coupling $g_x$ and the mixing angle $\alpha$.
%%%
As discussed in Sec.~\ref{Sec:Mass_spectrum_of_gauge_mediators}, in the light $Z'$ regime the $\rho$ parameter constraint requires the mixing angle $\alpha$ to be small. In this limit, we can safely approximate  $\cos\alpha \approx 1$ and $\sin \alpha \simeq 2 (\cos{\theta_{W}}/ g) X_{\Phi}g_x$.
%%%
%%%
This makes the BSM contribution to the $Z$ boson coupling proportional to $g_x^2$. Hence, this additional contribution can be reasonably neglected in comparison to the SM. 
%%%
%%%
This implies that the $Z$ boson couplings to fermions remain identical to their SM values, given by 
\begin{eqnarray}
g_{\psi_{\mathtt{L}}}^{z} \simeq (g_{\psi_{\mathtt{L}}}^{z})^{\text{SM}} = T_{\psi_{\mathtt{L}}}^{3} - Q_{\psi}\sin^{2}\theta_{W}, \, \quad
g_{\psi_{\mathtt{R}}}^{z} \simeq (g_{\psi_{\mathtt{R}}}^{z})^{\text{SM}} = - Q_{\psi}\sin^{2}\theta_{W}.
\end{eqnarray}
%%%%%%%%%%
Consequently, the neutral current weak interactions also remain identical to those in the SM in this regime.
%%%%%%%%%%
Having discussed the $Z$ boson couplings, we now turn to the $Z'$ couplings.
%%%%%%%
The $Z'$ couplings to SM fermions are given in the  Eqs.~\eqref{SM Couplings3} and \eqref{SM Couplings4}. Again, in the limit required to satisfy the $\rho$ parameter constraint, 
these couplings reduce to,
%%%%%%%%%%%
\begin{equation}\label{EQ:Zp Couplings}
g_{\psi_{_{\mathtt{L}}}}^{z'} \simeq  \left[2 X_{\Phi} \Big(T_{\psi_{\mathtt{L}}}^{3} - Q_{\psi} \sin^{2} \theta_{W} \Big)+ X_{\psi_{_{\mathtt{L}}}} \right]g_x\,,\,~~
 g_{\psi_{_{\mathtt{R}}}}^{z'} \simeq \left[   2 X_{\Phi}\Big( - Q_{\psi} \sin^{2} \theta_{W} \Big) + X_{\psi_{_{\mathtt{R}}}} \right] g_x\,.
\end{equation}
%%%%%%%%%%%
%%
This simplifies the structure of the couplings, with $g_x$ appearing as an overall factor. Moreover, since the Higgs doublet $\Phi$, possessing the larger VEV, effectively induces mixing between the $Z$ and $Z'$ bosons; consequently, a factor proportional to its charge $X_{\Phi}$ appears in Eq.~\eqref{EQ:Zp Couplings}.
This coupling structure has important phenomenological implications: 
%%%%%%%%%%%%%%%%
\begin{itemize}
    \item In light $Z'$ scenarios, the couplings of the $Z'$ to charged leptons are stringently constrained by a variety of experiments, including M\o ller scattering and measurements of $(g-2)_{e/\mu}$ \cite{SLACE158:2005uay,Jegerlehner:2009ry,Parker_2018,Rbegm2,Aliberti:2025beg,Athron:2025ets,DeRomeri:2025nkx}. The couplings to neutrinos are also constrained by both electron and nucleon scattering experiments 
\cite{AristizabalSierra:2018eqm,Majumdar:2021vdw,Coloma:2022avw,A:2022acy, Majumdar:2024dms,Blanco-Mas:2024ale,DeRomeri:2024iaw,Chattaraj:2025rtj}.
%%%%
Such constraints can be alleviated if the $Z'$ couplings to electrons or neutrinos are sufficiently suppressed or vanishes. The structure of the couplings in Eq.~\eqref{EQ:Zp Couplings} allows for this possibility. 
Requiring that the $Z'$ couplings to both the left- and right-handed components of a SM fermion $\psi$, with electric charge $Q_\psi$ and weak isospin $T_{\psi_{\mathtt{L}}}^{3}$, vanish simultaneously imposes the following conditions,
\begin{equation}\label{Eq:Zero_coupling_charges}
\frac{X_{\psi_{{\mathtt{L}}}}}{X_{\psi_{{\mathtt{R}}}}} = 1 - \frac{T_{\psi_{\mathtt{L}}}^{3}}{Q_{\psi} \sin^{2} \theta_{W}}\,,
\qquad
X_{\Phi} = \frac{X_{\psi_{{\mathtt{R}}}}}{2Q_{\psi} \sin^{2} \theta_{W}}\,.
\end{equation}
Thus, experimental constraints arising from processes involving any given SM fermion can be relaxed in this manner, as demonstrated for neutrinos in Sec.~\ref{subsec:Neutrino_couling}.
%%%%%%%%%%%%%%%
\item Another interesting aspect of this setup is the structure of the vector and axial vector couplings.
%%%%%%
Using Eq.~\eqref{Eq:AVCouplings_Zp} and Eq.~\eqref{EQ:Zp Couplings}, the vector and axial vector coupling of fermion with $Z'$ can be written as,
%%%%%%%%%%%%%%%%%%%%%%%
\begin{equation}\label{Eq:VA_coup}
    C^{\psi}_{V} \simeq \left[  X_{\Phi} (T_{\psi_{_{\mathtt{L}}}}^{3} - 2Q_{\psi} \sin^{2}{\theta_{W}}) + C^{\psi}_{V(0)}\right]g_x,~~~~ C^{\psi}_{A} \simeq \left[ X_{\Phi} T_{\psi_{_{\mathtt{L}}}}^{3} + C^{\psi}_{A(0)} \right]g_x\,.
\end{equation}
%%%%%%%%%%%%%%%%%%%%%%%

Here, $C^{\psi}_{V(0)/A(0)} = (X_{\psi_{\mathtt{L}}} \pm X_{\psi_{\mathtt{R}}})/2$ denote the vector and axial-vector charges of the $Z'$ to the fermion $\psi$ in the limit, $X_{\Phi}=0$.
%%%%
The $U(1)_X$ charge of Higgs with larger VEV, $X_{\Phi}$ (see Sec. \ref{Sec:Mass_spectrum_of_gauge_mediators}) appearing in Eq.~\eqref{Eq:VA_coup} plays an important role in determining the nature of the vector and axial-vector couplings.
%%%%%%%%%%%%%%%%%

\item The $U(1)_X$ charge of the Higgs is not a free parameter. As discussed in Sec.~\ref{Sec:anomaly_cancellations}, it is constrained by the requirement that Yukawa interactions that generate SM fermion masses should be gauge invariant. If the Higgs doublet with larger VEV $\Phi$ appearing in Eq.~\eqref{Eq:VA_coup} also generates the mass of the fermion $\psi$ via Yukawa interaction, its $U(1)_X$ charge is given by $X_{\Phi} = -(X_{\psi_{\mathtt{L}}} - X_{\psi_{\mathtt{R}}})/2 T_{\psi_{_{\mathtt{L}}}}^{3} = - C^{\psi}_{A(0)}/T_{\psi_{_{\mathtt{L}}}}^{3}$, see Eq. \eqref{Eq:Higgs_Charge} in Sec. \ref{Sec:anomaly_cancellations}.
%%%%
In vector models, where the left and right handed fermions carry identical charges ($X_{\psi_{{\mathtt{L}}}} = X_{\psi_{{\mathtt{R}}}} = X$), the axial vector charge vanishes, $C^{\psi}_{A(0)}=0$.
Consequently, the Higgs charge under $U(1)_X$ also becomes zero, $X_\Phi = 0$.
In such models, the vector coupling in the fermion gauge basis is given by $``Xg_x$'', while the axial vector coupling vanishes identically.
Consequently, the corresponding gauge boson exhibits purely vector couplings to SM fermions.
%%%%
Notable examples include $B-L$, $B - 3L_i$, $B_i - 3L_j$ and their possible linear combinations, where $i, j = 1, 2, 3$ label the fermion generations \footnote{In the case of generation-specific vector symmetries, such as $B_i - 3L_j$, mixed vector axial-vector couplings can appear in the fermion mass basis; however, this typically comes at the expense of inducing fermion mixing.}.
%%%%%%%%%%%%%%%%%%%%%%%%
%%%%%%%%%%%%%%%%%%%%%%%%
\item In case of chiral $U(1)_X$ models, if the same Higgs doublet $\Phi$ that acquires the larger VEV, also generates the mass of fermion $\psi$ through Yukawa interactions, then its $U(1)_X$ charge must satisfy the same relation as before, $X_{\Phi} = - C^{\psi}_{A(0)}/T_{\psi_{_{\mathtt{L}}}}^{3}$. 
%%%%
Substituting this relation into Eq.~\eqref{Eq:VA_coup}, it can be seen 
that the corresponding axial-vector coupling vanishes.
%%%%
As a consequence, flavor universal chiral models, such as linear combinations of hypercharge and $B-L$, effectively reduce to purely vector like theories in the light $Z'$ regime.
%%%%
Hence, in the light $Z'$ regime and within this framework, realizing mixed vector axial-vector or purely axial-vector couplings of the $Z'$ to SM fermions is difficult from the perspectives of both gauge anomaly cancellation and the SM fermion mass generation mechanism (see Sec.~\ref{Sec:anomaly_cancellations}).
%%%%%%%%%%%%%%%%%%%%%%%%%%%%%%%%%%%
%%%%%%%%%%%%%%%%%%%%%%%%%%%%%%%%%%%
\item For chiral $U(1)_X$ models, axial-vector couplings can arise if at least two Higgs doublets are present, such that the Higgs charge appearing in Eq.~\eqref{Eq:VA_coup} differs from the Higgs charge required for fermion mass generation through Yukawa interactions. In the two Higgs doublet setup discussed in Sec.~\ref{Sec:Mass_spectrum_of_gauge_mediators}, the Higgs $\Phi$ acquires the larger VEV, while $\varphi$ has a smaller VEV. Since Higgs with larger VEV effectively induces mixing between $Z$ and $Z'$, hence its charge $X_{\Phi}$ appears in Eq.~\eqref{Eq:VA_coup}.  
%%%
In flavor specific chiral $U(1)_X$ models, it is possible that $\varphi$ 
generates the masses of the first two generations, while $\Phi$ is 
responsible for the mass generation of the third generation \footnote{
This occurs when the first two generations of SM fermions carry the 
same $U(1)_X$ charge, while the third generation carries a different 
$U(1)_X$ charge.}. Hence, the Higgs charge entering Eq.~\eqref{Eq:VA_coup} differs from that involved in the Yukawa mass generation for the lighter fermions. Consequently, mixed vector--axial-vector or purely axial-vector couplings can arise for these fermions, as discussed in Sec.~\ref{subSec:Pure Vector and Axial Vector Couplings}.
\end{itemize}

%%%%%%%%%%%%%%%%%%%%%%%%%%%%%%%%%%
%%%%%%%%%%%%%%%%%%%%%%%%%%%%%%%%%%
%%%%%%%%%%%%%%%%%%%%%%%%%%%%%%%%%%
\subsection{Heavy $Z'$ scenario: $\mathbf{M_{Z'} > M_{Z}}$ }
%%%%%%
In the heavy $Z'$ case, the couplings of fermions to the $Z$ and $Z'$ bosons can be obtained from Eq.~\eqref{SM Couplings} by making the substitution $\alpha \to -\alpha$. The $Z$ boson couplings receive both SM and BSM contributions due to mixing, with the latter being suppressed by the mixing angle $\alpha$. To ensure that the fermion couplings to the $Z$ boson remain consistent with their SM values for a given $g_x$, the $Z'$ mass must be sufficiently large so that the mixing angle $\alpha$ remains small, as seen from Fig. \ref{FiG:Alpha_rho_Heavy_Zp}.
%%%
The $Z'$ couplings to fermions also contain a SM component, which is suppressed by the mixing angle $\alpha$.
%%%%
In this scenario, no simple approximate expression for the $Z'$ couplings to fermions, analogous to Eq.~\eqref{EQ:Zp Couplings}, is available, as $\alpha$ remains function of $g_x$, $M_{Z'}$, and $X_{\Phi}$, as shown in Fig. \ref{FiG:Alpha_rho_Heavy_Zp}. 
%%%%%%  
In this case, purely vector couplings arise when $X_{\Phi} = 0$ and $X_{\psi_{\mathtt{L}}} = X_{\psi_{\mathtt{R}}} = X$. 
They can also be realized in the limit where $M_{Z'}$ is sufficiently heavy, such that the SM contributions to the effective couplings can be neglected and $X_{\psi_{\mathtt{L}}} = X_{\psi_{\mathtt{R}}}$.
%%%%%%%%%%%%  
Mixed vector and axial-vector couplings can be realized if the fermion charge assignments are chiral. However, a purely axial-vector coupling is difficult to realize within this setup, as $\alpha$ remains function of $g_x$, $M_{Z'}$, and $X_{\Phi}$.
%%%%%%%%%%%%  

Notice that in this section the gauge boson masses, their mixing, and the fermionic couplings were discussed by treating the Higgs $U(1)_X$ charges $X_{\Phi}$ and $X_{\varphi}$, as well as the fermion $U(1)_X$ charges $X_{\psi}$, as free parameters. 
However, these charge assignments are not arbitrary. They are constrained both by the gauge anomaly cancellation conditions and by the requirements imposed by the Yukawa couplings. We analyse these constraints in detail in the next section.
%%%%%%%%%%%%%%%%%%%%%%%%%%%%%%%%%%%%%%%%%%%%%%%%%%%%%%%%%%%%%%%%%%%%%%%%%%%%%%%%%%%%%%
\section{Anomaly Cancellation}\label{Sec:anomaly_cancellations}
%%%%%%%%%%%%%%%%%%%%%%%%%%%%%%%%%%%%%%%%%%%%%%%%%%%%%%%%%%%%%%%%%%%%%%%%%%%%%%%%%%%%%%
In this section, we discuss gauge anomalies arising in the presence of the additional $U(1)_X$ symmetry. To preserve the consistency of the theory, such as its unitarity and renormalizability, all gauge anomalies must cancel \cite{Adler:1969gk,Bardeen:1969md,Bell:1969ts,Delbourgo:1972xb,Alvarez-Gaume:1983ihn, Witten:1982fp}.
In addition to the SM gauge anomaly cancellation conditions, the extended gauge group $SM \otimes U(1)_X$ generally gives rise to several triangle anomalies involving the $U(1)_X$ current, as listed below \cite{Prajapati:2024wuu},
%%
%%
%%%%%%
\begin{subequations}\label{Eq:U(1)x_anomaly_cancelation_charges}
\begin{align}
&[SU(3)_{C}]^2[U(1)_{X}]= \sum_{i=1}^{3}(2 X_{Q^{^{i}}} -  X_{u_{_{\mathtt{R}}}^{^{i}}}-X_{d_{_{\mathtt{R}}}^{^{i}}}) \label{Eq:anoUx1} = 0\,, 
\\&[SU(2)_{\mathtt{L}}]^2[U(1)_{X}]= \sum_{i=1}^{3}( X_{L^{^{i}}} +3 X_{Q^{^{i}}}) = 0\label{Eq:anoUx2}\,,
\\&[U(1)_{Y}]^2 [U(1)_{X}]= \sum_{i=1}^{3} ( X_{L^{^{i}}} + \frac{1}{3}  X_{Q^{^{i}}} -2  X_{e_{_{\mathtt{R}}}^{^{i}}} -\frac{8}{3}X_{u_{_{\mathtt{R}}}^{^{i}}}-\frac{2}{3}  X_{d_{_{\mathtt{R}}}^{^{i}}} ) = 0\label{Eq:anoUx3}\,,  
\\&[U(1)_{Y}] [U(1)_{X}]^2=  \sum_{i=1}^{3} \bigl\{   (X_{Q^{^{i}}})^{2}-(X_{L^{^{i}}})^{2}  + (X_{e_{_{\mathtt{R}}}^{^{i}}})^2 -2 (X_{u_{_{\mathtt{R}}}^{^{i}}})^2 + (X_{d_{_{\mathtt{R}}}^{^{i}}})^2  \bigl\} = 0 \label{Eq:anoUx4} \,,
\\ & [U(1)_{X}]^3= \sum_{i=1}^{3} \left[ 2(X_{L^{^{i}}})^{3}  +6 (X_{Q^{^{i}}})^{3}   - (X_{e_{_{\mathtt{R}}}^{^{i}}})^{3} -3 \bigl\{ (X_{u_{_{\mathtt{R}}}^{^{i}}})^{3}  + (X_{d_{_{\mathtt{R}}}^{^{i}}})^{3} \bigl\} \right]  - \sum_{i=1}^{3} (X_{\nu_{_{\mathtt{R}}}^{i}})^{3} = 0 \label{Eq:anoUx5} \,,
\\&[G]^2[U(1)_{X}]= \sum_{i=1}^{3} \bigl\{ 2X_{L^{^{i}}}  + 6 X_{Q^{^{i}}} -X_{e_{_{\mathtt{R}}}^{^{i}}}  -3\left( X_{u_{_{\mathtt{R}}}^{^{i}}} + X_{d_{_{\mathtt{R}}}^{^{i}}}\right) \bigl\}  - \sum_{i=1}^{3} X_{\nu_{_{\mathtt{R}}}^{i}} = 0 \label{Eq:anoUx6}\,. 
\end{align}
\end{subequations}
%%%%
Here, $X_{\psi}$ denote the $U(1)_{X}$ charge of the fermion $\psi$. The indices $i$ and $j$ label the fermion generations, with $i,j = 1,2,3$. The symbols $L = (\nu_{\mathtt{L}}, e_{\mathtt{L}})^{T}$ and $Q = (u_{\mathtt{L}},d_{\mathtt{L}})^{T}$ represent the SM lepton and quark $SU(2)_{L}$ doublets, while $l=e_{\mathtt{R}}$ and $q=(u_{\mathtt{R}},d_{\mathtt{R}})$ denote the SM lepton and quark $SU(2)_{L}$ singlets, respectively.
To ensure the gauge anomaly cancellation, all six conditions in Eq.~\eqref{Eq:U(1)x_anomaly_cancelation_charges}, together with the anomalies involving purely SM currents, must vanish.
To cancel the gauge anomalies, we also introduce three SM singlet right-handed neutrinos, 
$\nu_{\mathtt{R}}^{i}$ ($i=1,2,3$), each carrying a $U(1)_X$ charge 
$X_{\nu_{\mathtt{R}}^{i}}$.
Following the discussion in Sec. \ref{Sec:Interactions}, we consider flavor specific fermion $U(1)_X$ charges, allowing for mixed vector-axial vector couplings in both heavy and light $Z'$ scenarios.
%%%%
%%%%
For flavor specific scenarios, the $U(1)_X$ charges of the three generations can, in principle, be different. 
%%%%
However, due to the relatively stringent experimental limits on lepton flavor universality violation in the first two generations, arising from measurements such as $R_K$ and $R_{K^*}$ \cite{LHCb:2022qnv,LHCb:2022vje}, we assume identical $U(1)_X$ charge assignments for the first two generations.
%%%%
%%%%
In this setup, two generations of fermions carry the same $U(1)_X$ charges, while the third generation carries different charges.
%%%%

While solving Eqs.~(\ref{Eq:anoUx1}--\ref{Eq:anoUx6}), we also consider the SM fermions to be chiral under $U(1)_X$ in general.
The chiral nature of the SM fermions under $U(1)_X$ implies one should be careful in writing the Yukawa interactions.
The Yukawa interactions are,
\begin{equation}
   - \mathcal{L_{Y}} = Y_{e}^{ij}\overline{L^{i}} H e_{_{\mathtt{R}}}^{j} +Y_{u}^{ij} \overline{Q^{i}} \tilde{H} u_{_{\mathtt{R}}}^{j}  + Y_{d}^{ij} \overline{Q^{i}} H d_{_{\mathtt{R}}}^{j} + \text{h.c.}\,.  
\end{equation}
Here $H$ is the scalar doublet. In the two Higgs doublet setup considered in this work, $H$ represent either $\Phi$ or $\varphi$. Where, $\Phi (\varphi)$ is the Higgs doublet with the larger (smaller) VEV (see Sec, \ref{Sec:Mass_spectrum_of_gauge_mediators}). 
Requiring the corresponding Yukawa couplings to be gauge invariant under $U(1)_X$ imposes the following constraints on the $U(1)_X$ charges of the Higgs doublet(s),
\begin{equation}\label{Eq:Higgs_Charge}
    X_{H} = -\frac{X_{\psi_{\mathtt{L}}}^{i} - X_{\psi_{\mathtt{R}}}^{j}}{2 T_{3}^{\psi}}\,.
\end{equation}
%%%
As discussed in Sec.~\ref{Sec:Interactions}, if the Higgs doublet that 
acquires the larger VEV also generates fermion masses through Yukawa interactions satisfying Eq.~\eqref{Eq:Higgs_Charge} ($X_{H}=X_{\Phi}$), the axial-vector current associated with the $Z'$ interaction vanishes 
in the light $Z'$ limit, see Eq.~\eqref{Eq:VA_coup} and discussion afterwards.
%%%
Consequently, if fermion mass generation arises solely from a single Higgs doublet ($\Phi$), the axial-vector coupling in light $Z'$ scenarios vanishes. 
%%%
However, in the two Higgs doublet scenario considered in this work, the Higgs with the larger VEV ($\Phi$) can generate third-generation masses, while a second Higgs with a smaller VEV ($\varphi$) generates those of the first and second generations. In such a case, mixed vector axial-vector or even purely axial-vector couplings can be realized for the lighter fermions.
%%%
%%%

Motivated by these considerations, we discuss anomaly free solutions for two cases: those where fermion masses arise from a single Higgs doublet, and those requiring contributions from both. In what follows, we examine each case in detail.
%%%%%%%%%%%%
%%%%%%%%%%%%
%%%%%%%%%%%%
\subsection{ Scenario (I) : Single Higgs generates masses of all SM fermions}\label{Subsec:SI_Anomaly}
%%%%
We begin by examining the Scenario~(I) in which the Higgs doublet $\Phi$ alone accounts for the mass generation of all the SM fermions. 
%%%%
As mentioned earlier, we also assume that all SM fermions are charged under $U(1)_X$, with two generations sharing identical charges and one generation carrying distinct charge. To cancel the resulting gauge anomalies, three SM singlet right handed neutrinos ($\nu_{\mathtt{R}}^{i},\nu_{\mathtt{R}}^{3}$) are introduced.
%%% 
In our convention, we choose the ordering such that the first and second generations carry identical $U(1)_{X}$ charges ($i=1,2$), which differ from those of the third generation. 
%%%
However, since the gauge anomaly cancellation conditions depend only on sums (or sums of cubes/squares) of the charges over all generation of fermions, any permutation of the generation labels leaves the anomaly cancellation unchanged.
%%%
In particular, the $U(1)_X$ charges of fermions of the same SM representation, such as the quark doublets $Q^i$ ($i=1,2$) and $Q^3$, or the lepton doublets $L^i$ and $L^3$, can be freely interchanged without affecting the anomaly free character of the charge assignment.
%%%
%%%

Since a single Higgs doublet $\Phi$ is responsible for generating all SM fermion masses, the Yukawa interactions impose the following constraints on the allowed 
$U(1)_X$ charge assignments,
%%%
\begin{equation}\label{Eq:S1_Yukawa}
     X_{\Phi} = X_{L^{i(3)}} - X_{e_{_{\mathtt{R}}}^{j(3)}} = X_{u_{_{\mathtt{R}}}^{j(3)}} - X_{Q^{i(3)}} = X_{Q^{i(3)}} - X_{d_{_{\mathtt{R}}}^{j(3)}}\,.
\end{equation}
Here $i,j =1,2$ represents the first and second generations of fermions.
%%%
%%%%
\begin{table}[ht]
\centering
%\captionsetup{justification=centering}
\renewcommand{\arraystretch}{1.5}
\setlength{\tabcolsep}{-0.5pt} % reduce horizontal padding
\small
\begin{tabularx}{\textwidth}{|c c c|c c c|}
\hline
\centering
Fields &~ $SU(3)_C\!\otimes\!SU(2)_L\!\otimes\!U(1)_Y$ & $U(1)_X$ &
Fields &~ $SU(3)_C\!\otimes\!SU(2)_L\!\otimes\!U(1)_Y$ & $U(1)_X$ \\ 
\hline
$Q^{i}$ & $(3,2,1/3)$ & $\tfrac{-1}{6}(2l+3y+4l_{3}-3a)$ & $L^{i}$ & $(1,2,-1)$ & $l$ \\ 
$u_{\mathtt{R}}^{i}$ & $(3,1,4/3)$ & $\tfrac{-1}{6}(2l+3y+10l_{3}-9a)$ & $e_{\mathtt{R}}^{i}$ & $(1,1,-2)$ & $l+l_{3}-a$ \\ 
$d_{\mathtt{R}}^{i}$ & $(3,1,-2/3)$ & $\tfrac{-1}{6}(2l+3y-2l_{3}+3a)$ & $\nu_{\mathtt{R}}^{i}$ & $(1,1,0)$ & $l-l_{3}+a$ \\ 
\hline
$Q^{3}$ & $(3,2,1/3)$ & $y+ l_{3} - a$ & $L^{3}$ & $(1,2,-1)$ & $l_{3}$ \\ 
$u_{\mathtt{R}}^{3}$ & $(3,1,4/3)$ & $y$ & $e_{\mathtt{R}}^{3}$ & $(1,1,-2)$ & $2l_{3}-a$ \\ 
$d_{\mathtt{R}}^{3}$ & $(3,1,-2/3)$ & $y+2l_{3}-2a$ & $\nu_{\mathtt{R}}^{3}$ & $(1,1,0)$ & $a$ \\
\hline
$\Phi$ & $(1,2,1)$ & $a-l_{3}$ & & & \\
\hline
\end{tabularx}
\caption{Charge assignments of SM and BSM fields under the SM gauge group and $U(1)_X$. A single Higgs doublet, $\Phi$, is sufficient to generate the masses of all SM fermions. Furthermore, the charge of the singlet scalar $\chi_0$ needed to generate mass of $Z'$ remains completely unconstrained. }
\label{parametertable}
\end{table}
%%%%%%%%
Solving the gauge anomaly cancellation conditions in Eq.~\eqref{Eq:U(1)x_anomaly_cancelation_charges} together with the Yukawa constraints in Eq.~\eqref{Eq:S1_Yukawa} yields the charge assignments shown in Table~\ref{parametertable}. The solution is parameterised by four independent parameters $(a,l,l_3,y)$, and all remaining charges are determined in terms of these.
The charge assignments are flavor specific and can therefore induce FCNCs in both the quark and lepton sectors.
The $U(1)_X$ charges of the lepton doublets are $l$ and $l_3$. These distinct charges make the lepton sector generation-specific, and in the limit $l_3 \to l$, the lepton sector becomes flavor universal.
Note that the axial-vector charges $C^{\psi}_{A(0)} = 
(X_{\psi_{\mathtt{L}}} - X_{\psi_{\mathtt{R}}})/2$,  are proportional to $a - l_3$ for all fermions. Therefore, in the 
limit $a \to l_3\, (\alpha \to 0)$, the model reduces to a generation-specific vector model in the fermion gauge basis (see Eq. \eqref{Eq:Vector_axialVector_Coupling}).
%%%%%%%%%%
Furthermore, this charge assignment also allows the Yukawa interaction $\overline{L^{i(3)}} \tilde{\Phi} \nu_{\mathtt{R}}^{j(3)}$. Consequently, the masses of all SM fermions, including the neutrinos, can be generated within this setup. 
%%%%%%%%%%
The generation of quark and lepton mixing is further discussed in Sec.~\ref{Sec:Quark_mixing} and Sec.~\ref{Sec:Lep_mix}, respectively.
%%%%%%%%%%
%%%%%%%%%%
%%%%%%%%%%
%%%%%%%%%%
%%%%%%%%%%
\subsection{ Scenario (II) : Two Higgses generate masses of SM fermions}\label{Subsec:SII_Anomaly}
%%%%%%%%%%
%%%%%%%%%%
%%%%%%%%%%
%%%%%%%%%%
We now consider a framework with two Higgs doublets, $\Phi$ and $\varphi$, generating the SM fermion masses through Yukawa interactions. We assume that $\varphi$ couples to the first two generations, which share identical charges, while $\Phi$ generates the masses of the third generation \footnote{As noted earlier, any alternative ordering of fermion generations can be adopted without affecting the anomaly cancellation conditions.}. Consequently, the $U(1)_X$ charges must satisfy,
%%%
\begin{equation}\label{Eq:S2_Yukawa}
    \begin{split}
   & X_{\varphi} = X_{L^{i}} - X_{e_{_{\mathtt{R}}}^{j}} = X_{u_{_{\mathtt{R}}}^{j}} - X_{Q^{i}} = X_{Q^{i}} - X_{d_{_{\mathtt{R}}}^{j}}\,,\\
& X_{\Phi} = X_{L^{3}} - X_{e_{_{\mathtt{R}}}^{3}} = X_{u_{_{\mathtt{R}}}^{3}} - X_{Q^{3}} = X_{Q^{3}} - X_{d_{_{\mathtt{R}}}^{3}}\,.            
    \end{split}
\end{equation}
%%%
Solving Eq.~\eqref{Eq:U(1)x_anomaly_cancelation_charges} together with the constraints from Eq.~\eqref{Eq:S2_Yukawa} determines the charge assignments displayed in Table~\ref{parametertable_Two}.
%%%%
%%%%
\begin{table}[ht]
\centering
\renewcommand{\arraystretch}{1.5}
\setlength{\tabcolsep}{5pt} % balanced horizontal padding
\small
\begin{tabular}{|c c c|c c c|}
\hline
Fields & $SU(3)_C\!\otimes\!SU(2)_L\!\otimes\!U(1)_Y$ & $U(1)_X$ &
Fields & $SU(3)_C\!\otimes\!SU(2)_L\!\otimes\!U(1)_Y$ & $U(1)_X$ \\ 
\hline
$Q^{i}$ & $(3,2,1/3)$ & $-\frac{l}{3}$ & $L^{i}$ & $(1,2,-1)$ & $l$ \\ 
$u_{\mathtt{R}}^{i}$ & $(3,1,4/3)$ & $\frac{2l}{3} - x$ & $e_{\mathtt{R}}^{i}$ & $(1,1,-2)$ & $x$ \\ 
$d_{\mathtt{R}}^{i}$ & $(3,1,-2/3)$ & $-\frac{4l}{3} + x$ & $\nu_{\mathtt{R}}^{i}$ & $(1,1,0)$ & $2l - x$ \\
\hline
$Q^{3}$ & $(3,2,1/3)$ & $-\frac{l_3}{3}$ & $L^{3}$ & $(1,2,-1)$ & $l_{3}$ \\ 
$u_{\mathtt{R}}^{3}$ & $(3,1,4/3)$ & $a - \frac{4l_3}{3}$ & $e_{\mathtt{R}}^{3}$ & $(1,1,-2)$ & $2l_{3} - a$ \\ 
$d_{\mathtt{R}}^{3}$ & $(3,1,-2/3)$ & $-a + \frac{2l_3}{3}$ & $\nu_{\mathtt{R}}^{3}$ & $(1,1,0)$ & $a$ \\ 
\hline
$\varphi$ & $(1,2,1)$ & $l - x$ & $\chi_0$ & $(1,1,0)$ & $a + x-(l + l_3)  $ \\
$\Phi$ & $(1,2,1)$ & $a - l_{3}$ & & & \\
\hline
\end{tabular}
\caption{Charge assignments of SM and BSM fields under the SM gauge group and $U(1)_X$. The scalar $\Phi$ generates masses for the third generation, while $\varphi$ generates those for the remaining two. The singlet $\chi_0$ allows the $Z'$ mass to vary freely, and its $U(1)_X$ charge is chosen such that no massless Goldstone bosons appear in the model.}
\label{parametertable_Two}
\end{table}
%%%%% 
The $U(1)_X$ charge assignments are parametrized in terms of the lepton charges $(l, l_3, x, a)$ and quark charges are represented in terms of them.
The scalar doublets $\Phi$ and $\varphi$, responsible for generating the SM fermion masses, carry charges as listed in Table~\ref{parametertable_Two}. 
As discussed in Sec.~\ref{Sec:Mass_spectrum_of_gauge_mediators}, additional scalar singlets are required to freely vary the $Z'$ mass. Accordingly, we introduce an SM singlet scalar $\chi_0$, which couples to the scalar doublets through cubic interactions and thereby avoids the appearance of unwanted massless Goldstone modes.
The fermionic charge assignments are chiral, leading to axial-vector charges $C^{\psi}_{A(0)} = (X_{\psi_{\mathtt{L}}} - X_{\psi_{\mathtt{R}}})/2$,  proportional to $(a - l_3)$ for the third generation and $(l - x)$ for the other two generations.
In the limit $a \to l_3$, the $U(1)_X$ charges of the third generation fermions reduce to those of the $B-L$ model, up to a free parameter $l_3$. Similarly, when $x \to l$, the charge assignments for the other two generations also reproduce the $B-L$ structure with a free parameter $l$.
Again, this charge assignment permits the Yukawa interaction $\overline{L^{i(3)}} \tilde{\varphi}(\tilde{\Phi}) \nu_{\mathtt{R}}^{j(3)}$. Consequently, the masses of all SM fermions, including neutrinos, can be generated within this framework.

In the following sections, we examine how quark and lepton mixing arise from the two general anomaly-free $U(1)_X$ charge assignments introduced above.

%%%%%%%%%%%%%%%%%%%%%%%%%%%%%%%%%%%%%%%%%%%%%%%%%%%%%%%%%%%%%%%%%%%%%%%%%%%%%%%%%%%%%%%%%%%%%%%%%%%%%%%%%%%%%%%%%%%%%%%%%%%%%%%%% Fermionic mixing and neutrino nature
\section{Quark mixing pattern}\label{Sec:Quark_mixing}
%%%%%%%%%%%%%%%%%%%%%%%%%%%%%%%%%%%%%%%%%%%%%%%%%%%%%%%%%%%%%%%%%%%%%%%%%%%%%%%%%%%%%%%%%%%%%%%%%%%%%%%%%%%%%%%%%%%%%%%%%%%%%%%%%
In this section, we address the generation of quark mixing, i.e., the structure of the Cabibbo-Kobayashi-Maskawa (CKM) matrix~\cite{Cabibbo:1963yz,Kobayashi:1973fv}, in the context of the two anomaly-free charge assignments of Sec.~\ref{Sec:anomaly_cancellations}.
%%
% In this section, we discuss the generation of  quark mixing pattern, i.e. the Cabibbo-Kobayashi-Maskawa (CKM) matrix \cite{Cabibbo:1963yz,Kobayashi:1973fv}, for the two anomaly-free charge assignments introduced in Sec.~\ref{Sec:anomaly_cancellations}.
While Scenario~(I) can generate all fermion masses via a single Higgs doublet, mixing can be realized either by fixing specific free parameters within the $U(1)_X$ charges or by introducing an additional Higgs doublet. In contrast, Scenario~(II) utilizes two Higgs doublets from the outset. 
Accordingly, we classify these models based on whether a single Higgs is sufficient or if two are required to generate the observed quark masses and mixing.
% In this section we discuss how quark and lepton mixing can be generated for the two general anomaly free charge assignment discussed in Sec.~\ref{Sec:anomaly_cancellations}. 
% %%%
%%%%%%%%%%%%%%%%
\subsection{Models with single Higgs doublet}\label{Subsec:Qmix_1hdm}
%%%%%%%%%%%%%%%%
Here we discuss models featuring a single Higgs doublet. These models are based on Scenario~(I) (Sec.~\ref{Subsec:SI_Anomaly}), in which the Higgs field $\Phi$ generates masses for all SM fermions. 
%%%%%%%%%%%%%%%%
%%%
Table~\ref{parametertable} summarizes the particle content and their charges under the gauge group $\text{SM} \otimes U(1)_X$. 
The Yukawa terms responsible for generating the fermion masses can be written as,
%%%%%%%%%%%%
\begin{equation}
\begin{split}
-\mathcal{L}_{Y} = Y_{u}^{ij} \overline{Q^{i}} \tilde{\Phi} u_{\mathtt{R}}^{j} + Y_{u}^{33} \overline{Q^{3}} \tilde{\Phi} u_{\mathtt{R}}^{3} + Y_{d}^{ij} \overline{Q^{i}}\Phi d_{\mathtt{R}}^{j} +Y_{d}^{33} \overline{Q^{3}}\Phi d_{\mathtt{R}}^{3}   + Y_{e}^{ij} \overline{L^{i}} \Phi e_{\mathtt{R}}^{j}+ Y_{e}^{33} \overline{L^{3}} \Phi e_{\mathtt{R}}^{3}   \\
+ Y_{\nu}^{ij} \overline{L^{i}} \tilde{\Phi}    \nu_{\mathtt{R}}^{j}
  + Y_{\nu}^{33} \overline{L^{3}} \tilde{\Phi} \nu_{\mathtt{R}}^{3} + h.c.\,.
\end{split}
\end{equation}
%%%%%%%%%%%%%
Here, $\tilde{\Phi}  = i \sigma_{2} \Phi^{*}$, where $\sigma_{2}$ is the second Pauli matrix, and $i,j = 1,2$. The VEV of the Higgs field is given by $\langle \Phi \rangle = v_{\Phi}/\sqrt{2}$. After the spontaneous breaking of the $U(1)_{X}$ and electroweak symmetries, the mass matrices of both the quark and charged lepton sectors take the following form,
%%%
\begin{equation}\label{Eq:general_mass_marix_1}
   \mathcal{M} \sim \begin{bmatrix}
       \times & \times & 0\\
       \times & \times & 0 \\
       0 & 0 & \times
   \end{bmatrix}\,.
\end{equation}
%%%
Here $\times$ denotes matrix entries proportional to $v_{\Phi}$. With this fermion charge assignment, a single Higgs doublet suffices to generate fermion masses, but it cannot generate fermion mixing. For quark mixing, we need off diagonal terms in the mass matrix $\mathcal{M}$.
To generate the off-diagonal terms in the mass matrix of Eq.~\eqref{Eq:general_mass_marix_1}, the Higgs fields gets an additional constraints on its $U(1)_X$ charge assignments,
\begin{subequations}\label{Eq:Phi2_Charge_cross_term}
\begin{align}
    &\overline{Q^{3}} \Phi d_{\mathtt{R}}^{i} ~(\text{or}~ \overline{Q^{i}} \tilde{\Phi} u_{_\mathtt{R}}^{3} ) \longrightarrow X_{\Phi} = \frac{1}{6} (2 l + 4 l_3 +9y - 3 a )\,, \label{Eq:Phi2_Charge_cross_term1}\\
    & \overline{Q^{i}} \Phi d_{\mathtt{R}}^{3} ~(\text{or}~ \overline{Q^{3}} \tilde{\Phi} u_{\mathtt{R}}^{i} ) \longrightarrow X_{\Phi} = -\frac{1}{6} (2 l+16 l_3 +9y -15a)\label{Eq:Phi2_Charge_cross_term2}\,.  
\end{align}    
\end{subequations}
Note that the presence of a cross term in the down quark sector automatically ensures a corresponding cross term in the up quark sector.
Using charge assignment of Table. \ref{parametertable} and solving for two different choices of $X_{\Phi}$, as given in Eqs.~\eqref{Eq:Phi2_Charge_cross_term1} and \eqref{Eq:Phi2_Charge_cross_term2}, we obtain the same relation,
\begin{equation} \label{Eq:Y1}
y = a - \frac{2}{9}(l + 5l_3)\,.
\end{equation}
Hence, both choices of $X_{\Phi}$ yield identical fermionic charge assignments. Substituting the value of $y$ from Eq.~\eqref{Eq:Y1} into Table~\ref{parametertable} gives the charge assignment shown in Table \ref{Tab:parametertable1}.
%%%%%%%%%%%%%%%%%%%%%%%%%%%%%%%%%%
\begin{table}[ht]
\centering
%\captionsetup{justification=centering}
\renewcommand{\arraystretch}{1.5}
\setlength{\tabcolsep}{8pt} % control spacing between columns
\small
\begin{tabular}{|c c|c c|}
\hline
Fields &~~ $U(1)_X$ ~~& Fields & ~~$U(1)_X$~~ \\ 
\hline
$Q^{i}$ & $-\frac{1}{9}(2l+l_3)$ & $L^{i}$ & $l$ \\ 
$u_{\mathtt{R}}^{i}$ & $a-\frac{2}{9}(l+5l_3)$ & $e_{\mathtt{R}}^{i}$ & $l + l_{3} - a$ \\ 
$d_{\mathtt{R}}^{i}$ & $-a + \frac{2}{9}(4l_3 - l)$ & $\nu_{\mathtt{R}}^{i}$ & $a + l - l_{3}$ \\ 
\hline
$Q^{3}$ & $-\frac{1}{9}(2l+l_3)$ & $L^{3}$ & $l_{3}$ \\ 
$u_{\mathtt{R}}^{3}$ & $a-\frac{2}{9}(l+5l_3)$ & $e_{\mathtt{R}}^{3}$ & $2l_{3} - a$ \\ 
$d_{\mathtt{R}}^{3}$ & $-a + \frac{2}{9}(4l_3 - l)$ & $\nu_{\mathtt{R}}^{3}$ & $a$ \\ 
\hline
$\Phi$ & $a - l_{3}$ & & \\ 
\hline
\end{tabular}
\caption{Charges of SM and BSM particles under $U(1)_X$. Quark charges are flavor universal, and a single Higgs doublet $\Phi$ generates their masses and mixings. This Table can be obtained from Table \ref{parametertable} by putting $y = a - 2(l + 5l_3)/9\,$. Note that an additional singlet scalar $\chi_0$ is also needed to give mass to $Z'$, however  in this scenario its $U(1)_X$ charge is completely unconstrained. }
\label{Tab:parametertable1}
\end{table}
%%%%%%%%%
This particular choice of $y$ leads to flavor universal yet chiral quark charges. The lepton sector, being independent of $y$, remains unaffected.
%%%%%%%%

%%%%%%%%%%%%%%%%%%%%%%%%
\subsection{Models with two Higgs doublets}\label{Subsec:Qmix_2hdm}
%%%%%%%%%%%%%%%%%%%%%%%%
In this subsection, we consider models with two Higgs doublets.
%%%
For both anomaly-cancellation solutions discussed in 
Sec.~\ref{Sec:anomaly_cancellations} (Scenario~(I) and Scenario~(II)), 
quark mixing can be generated by a second Higgs doublet 
$\varphi$. We discuss these two scenarios in detail below.
%%%
\subsubsection{ Extensions of Scenario (I) }
%%%
We begin with extensions of Scenario (I), where the Higgs doublet $\Phi$ generates masses for all SM fermions. The resulting mass matrices are presented in Eq.~\eqref{Eq:general_mass_marix_1}.
%%%%
To induce off-diagonal entries in these mass matrices, and thereby generate fermion mixing, we introduce a second Higgs doublet $\varphi$. 
%%%
The constraints on the $U(1)_X$ charge of $\varphi$ required to produce these off-diagonal terms are obtained from Eq.~\eqref{Eq:Phi2_Charge_cross_term} by replacing $\Phi$ with $\varphi$.
%%%
In this case, to prevent massless Goldstone bosons, a cross term between the $SU(2)_L$ doublets, i.e., $\Phi^{\dagger} \varphi \chi_0$, is required. Here, $\chi_0$ with $U(1)_X$ charge $-(2l + 10l_3 + 9y_1 - 9a)/6$ allows such a term for both choices of $X_{\varphi}$ in Eq.~\eqref{Eq:Phi2_Charge_cross_term}. Hence, both choices are equivalent in this respect.
The resulting charges of particles are shown in Table \ref{Table:parametertable2}.
%%%%%%%
\begin{table}[ht]
\centering
\renewcommand{\arraystretch}{1.5}
\setlength{\tabcolsep}{10pt}
\small
\begin{tabular}{|c c|c c|}
\hline
Fields & $U(1)_X$ & Fields & $U(1)_X$ \\
\hline
$Q^{i}$ & $-\frac{1}{6}(2l + 4l_3 + 3y - 3a)$ & $L^{i}$ & $l$ \\
$u_{\mathtt{R}}^{i}$ & $-\frac{1}{6}(2l + 10l_3 + 3y - 9a)$ & $e_{\mathtt{R}}^{i}$ & $l + l_3 - a$ \\
$d_{\mathtt{R}}^{i}$ & $-\frac{1}{6}(2l + 3y - 2l_3 + 3a)$ & $\nu_{\mathtt{R}}^{i}$ & $a + l - l_3$ \\
\hline
$Q^{3}$ & $-a + l_3 + y$ & $L^3$ & $l_3$ \\
$u_{\mathtt{R}}^{3}$ & $y$ & $e_{\mathtt{R}}^{3}$ & $2l_3 - a$ \\
$d_{\mathtt{R}}^{3}$ & $2l_3 + y - 2a$ & $\nu_{\mathtt{R}}^{3}$ & $a$ \\
\hline
$\Phi$ & $a - l_3$ & $\chi_0$ & $-\frac{1}{6} (2l+10 l_3 +9 y_1 -9a)$ \\
$\varphi$ & $\frac{1}{6} (2l+4 l_3 + 9y -3a)$ or $-\frac{1}{6} (2 l+16 l_3 +9y -15a)$ &  & \\
\hline
\end{tabular}
\caption{Charges of SM and BSM particles under $U(1)_X$. While a single Higgs $\Phi$ is enough to generate masses of all SM fermions, the second Higgs doublet $\varphi$ is needed to generate the correct CKM mixing pattern. The charge of the $\chi_0$ is chosen to prevent the appearance of an unwanted massless Goldstone boson, see text.}
\label{Table:parametertable2}
\end{table}
%%%%%%%%%%%%%%
%

The mass matrices for the up and down quark sectors, corresponding to the choice of $X_{\varphi}$ given in Eq.~\eqref{Eq:Phi2_Charge_cross_term1}, are expressed as follows \footnote{For other choice of $X_{\varphi}$, matrix element proportional to $\varphi$ in up and down sector flips in position $\mathcal{M}^u \leftrightarrow \mathcal{M}^d$ .},
\begin{equation}\label{Eq:Mass_matrix_Up_down_two_higgs_case}
  \mathcal{M}^{u} \sim   \begin{bmatrix}
       \times & \times & \rd{\times}\\
       \times & \times & \rd{\times} \\
       0 & 0 & \times
   \end{bmatrix},~~ \mathcal{M}^{d} \sim \begin{bmatrix}
       \times & \times & 0\\
       \times & \times & 0 \\
       \rd{\times} & \rd{\times} & \times
   \end{bmatrix}\,.
\end{equation}
%%%%
Here, $\times$\, (\rd{$\times$}) denotes matrix elements proportional to $v_{\Phi}\,(v_{\varphi})$. The resulting mass matrices possess sufficient freedom to reproduce the CKM matrix. An advantage of employing two distinct Higgs doublets is that the off-diagonal elements in both the up and down sectors can be generated without reducing the number of free parameters in the fermion charge assignments.
%%%%%
%%%%%
\subsubsection{ Extensions of Scenario (II) }
%%%%%
%%%%%
We now turn our attention to models corresponding to Scenario~(II), as discussed in Sec.~\ref{Subsec:SII_Anomaly}. The anomaly-free $U(1)_X$ charge assignments for this case are summarized in Table~\ref{parametertable_Two}. This particular assignment requires two Higgs doublets to generate the masses for all SM fermions: the scalar doublet $\Phi$ generates masses for the third generation after SSB, while $\varphi$ accounts for the masses of the remaining two generations.
Additionally, as before,  the charges of the singlet scalar $\chi_0$ are always chosen such that the cubic term ($\Phi^\dagger \varphi \chi_0 \, + \, \rm{h.c.}$) is allowed, preventing the appearance of unwanted massless Goldstone bosons.
The Yukawa interactions responsible for generating the fermion masses are given by
%%%%%%%%
\begin{eqnarray}
-\mathcal{L}_{Y} &=& Y_{u}^{ij} \overline{Q^{i}} \tilde{\varphi} u_{\mathtt{R}}^{j} + Y_{u}^{33} \overline{Q_{3}} \tilde{\Phi} u_{\mathtt{R}}^{3} + Y_{d}^{ij} \overline{Q^{i}}\varphi d_{\mathtt{R}}^{j} +
Y_{d}^{33} \overline{Q^{3}}\Phi d_{\mathtt{R}}^{3} +
+ Y_{e}^{ij} \overline{L^{i}} \varphi e_{\mathtt{R}}^{j}
+ Y_{e}^{33} \overline{L^{3}} \Phi e_{\mathtt{R}}^{3}   \nonumber \\
&+& Y_{\nu}^{ij} \overline{L^{i}} \tilde{\varphi} \nu_{\mathtt{R}}^{j} + Y_{\nu}^{33} \overline{L^{3}} \tilde{\Phi} \nu_{\mathtt{R}}^{3}  +~ \text{h.c.}\,.
\end{eqnarray}
%%%%%%%%
Here, $\tilde{\Phi}(\tilde{\varphi}) = i\sigma_{2}\Phi^{*}(\varphi^{*})$, where $\sigma_{2}$ is the second Pauli matrix, and $i, j = 1, 2$. After SSB, the VEVs of $\Phi$ and $\varphi$ are given by $\langle \Phi \rangle = v_{\Phi}/\sqrt{2}$ and $\langle \varphi \rangle = v_{\varphi}/\sqrt{2}$, respectively. The corresponding mass matrices for the quark and lepton sectors are expressed as,
%%%
%%%
\begin{equation}\label{Eq:general_mass_marix_2}
   \mathcal{M} \sim \begin{bmatrix}
       \rd{\times} & \rd{\times} & 0\\
       \rd{\times} & \rd{\times} & 0 \\
       0 & 0 & \times
   \end{bmatrix}\,.
\end{equation}
%%%
%%%
Here matrix entries proportional to $v_{\Phi}$ and $v_{\varphi}$ is shown by $\times$ and $\red{\times}$ respectively. 
The charge assignment shown in Table~\ref{parametertable_Two} allows the generation of SM fermion masses but not their mixing. 
To generate the off-diagonal terms for quarks, the constraints on $U(1)_X$ charges of Higgs doublets are,
\begin{subequations}\label{Eq:Charge_Higgs_to_offdia_2}
\begin{align} 
    & \overline{Q^{3}} H d_{\mathtt{R}}^{j} \to X_{H} = \frac{1}{3} (4l - l_3) -x,~~~ \overline{Q^{3}} \tilde{H} u_{\mathtt{R}}^{j} \to X_{H} = \frac{1}{3} (2l + l_3) - x \,,  \label{Eq:Charge_Higgs_to_offdia_21}\\
    &  \overline{Q^{i}} H d_{\mathtt{R}}^{3} \to X_{H} = a - \frac{1}{3} ( l+ 2 l_3),~~~ \overline{Q^{i}} \tilde{H} u_{\mathtt{R}}^{3} \to X_{H} = a - \frac{1}{3} (4 l_3 - l)\, \label{Eq:Charge_Higgs_to_offdia_22}.
\end{align}    
\end{subequations}
To generate the CKM matrix, we begin by introducing off-diagonal entries in the down-quark sector. These off-diagonal elements can be generated in two possible ways, as described in Eqs.~\eqref{Eq:Charge_Higgs_to_offdia_21} and \eqref{Eq:Charge_Higgs_to_offdia_22}. Taking $X_H = X_\varphi$ leads to the following conditions for the two cases \footnote{An alternative choice, $X_H = X_\Phi$, yields the same conditions but with the two conditions of Eq.~\eqref{Eq:Off_dia_down_condition_l=l1}, mutually interchanged.},
\begin{equation} \label{Eq:Off_dia_down_condition_l=l1}
    \overline{Q^{3}} \varphi d_{\mathtt{R}}^{j} \to l= l_3\,\,,\,\,~~~~ \overline{Q^{i}} \varphi d_{\mathtt{R}}^{3} \to a = \frac{2}{3} (2l+l_3) - x\,.
\end{equation}
In the first case, generating the $\overline{Q^{3}} \varphi d_{\mathtt{R}}^{j}$ term requires $l = l_3$. In this limit, both the quark and lepton doublets become flavor universal, leading to the appearance of off-diagonal terms for the other fermions as well. This can be seen from Eq.~\eqref{Eq:Charge_Higgs_to_offdia_2} where putting $l = l_3$ gives a unique solution $X_H = l - x$ for all of the conditions. The resulting mass matrix can be written as,
%%
%%%
%%%
\begin{equation}\label{Eq:general_mass_marix_3}
   \mathcal{M}^{u/d/e/\nu} \sim \begin{bmatrix}
       \rd{\times} & \rd{\times}& \times\\
       \rd{\times} & \rd{\times} & \times \\
       \rd{\times} & \rd{\times} & \times
   \end{bmatrix}\,.
\end{equation}
%%%
%%%
Hence, both quark and lepton mixing matrices can be generated in this case.
%%%
The $U(1)_X$ charges of various particles in this case is given in Table \ref{Tab:parametertable5}.
%%%%%%
\begin{table}[ht]
\centering
%\captionsetup{justification=centering}
\renewcommand{\arraystretch}{1.5}
\setlength{\tabcolsep}{20pt} % adjusted horizontal padding
\small
\begin{tabular}{|c c|c c|}
\hline
Fields & $U(1)_X$ & Fields & $U(1)_X$ \\ 
\hline
$Q^{i}$ & $-\frac{l}{3}$ & $L^{i}$ & $l$ \\ 
$u_{\mathtt{R}}^{i}$ & $\frac{2l}{3} - x$ & $e_{\mathtt{R}}^{i}$ & $x$ \\ 
$d_{\mathtt{R}}^{i}$ & $-\frac{4l}{3} + x$ & $\nu_{\mathtt{R}}^{i}$ & $2l - x$ \\
\hline
$Q^{3}$ & $-\frac{l}{3}$ & $L^{3}$ & $l$ \\ 
$u_{\mathtt{R}}^{3}$ & $a - \frac{4l}{3}$ & $e_{\mathtt{R}}^{3}$ & $2l - a$ \\ 
$d_{\mathtt{R}}^{3}$ & $-a + \frac{2l}{3}$ & $\nu_{\mathtt{R}}^{3}$ & $a$ \\ 
\hline
$\varphi$ & $l - x$ & $\chi_0$ & $a + x - 2l $ \\
$\Phi$ & $a - l$ & & \\
\hline
\end{tabular}
\caption{Charges of SM and BSM particles under $U(1)_{X}$. The Higgs doublet $\Phi$ generates the masses of the third-generation fermions, while the masses of the first two generations, as well as the quark and lepton mixings, are generated by the second Higgs doublet $\varphi$. The charge of $\chi_0$ is chosen to ensure that no massless Goldstone modes remain in the model.}
\label{Tab:parametertable5}
\end{table}
%%%

In the second case, the $\overline{Q^i} H d_{\mathtt{R}}^{3}$ term gives condition, $a = 2(2l+l_3)/3 -x$. The $U(1)_X$ charges of particles for this case are shown in the Table \ref{Tab:parametertable6}. 
%%%%%%
\begin{table}[ht]
\centering
%\captionsetup{justification=centering}
\renewcommand{\arraystretch}{1.5}
\setlength{\tabcolsep}{20pt} % adjusted horizontal padding
\small
\begin{tabular}{|c c|c c|}
\hline
Fields & $U(1)_X$ & Fields & $U(1)_X$ \\ 
\hline
$Q^{i}$ & $-\frac{l}{3}$ & $L^{i}$ & $l$ \\ 
$u_{\mathtt{R}}^{i}$ & $\frac{2l}{3} - x$ & $e_{\mathtt{R}}^{i}$ & $x$ \\ 
$d_{\mathtt{R}}^{i}$ & $-\frac{4l}{3} + x$ & $\nu_{\mathtt{R}}^{i}$ & $2l - x$ \\ 
\hline
$Q^{3}$ & $-\frac{l_3}{3}$ & $L^{3}$ & $l_3$ \\ 
$u_{_\mathtt{R}}^{3}$ & $ - x + \frac{1}{3} (4l-2l_3)$ & $e_{\mathtt{R}}^{3}$ & $\frac{4}{3} (l_3 - l) + x$ \\ 
$d_{\mathtt{R}}^{3}$ & $ -\frac{4l}{3} + x$ & $\nu_{\mathtt{R}}^{3}$ & $\frac{2}{3}(2l + l_3) -x$ \\ 
\hline
$\varphi$ & $l - x$ & $\chi_0$ & $ \frac{1}{3} (l - l_3)$ \\
$\Phi$ & $ -x + \frac{1}{3}(4l - l_3)$ & & \\
\hline
\end{tabular}
\caption{Charges of SM and BSM particles under $U(1)_{X}$. The Higgs doublet $\Phi$ generates the third-generation fermion masses, while $\varphi$ generates the masses of the first two generations. The right-handed down-type quarks are flavor universal, generating the CKM matrix. Charge of $\chi_0$ is chosen to ensure absence of massless Goldstone modes in the model.}
\label{Tab:parametertable6}
\end{table}
%%%%%%%%%%
In this case, the charges of the right-handed down type quarks become flavor universal, leading to the generation of off diagonal terms in the down sector. However, no such terms arise in the lepton sector. The corresponding mass matrices are given as,
%%%
%%%
\begin{equation}\label{Eq:general_mass_marix_4}
   \mathcal{M}^{d} \sim \begin{bmatrix}
       \rd{\times} & \rd{\times} & \rd{\times}\\
       \rd{\times} & \rd{\times} & \rd{\times} \\
       \times & \times & \times
\end{bmatrix}\,,~~~~\,\mathcal{M}^{u/e/\nu} \sim \begin{bmatrix}
       \rd{\times} & \rd{\times} & 0\\
       \rd{\times} & \rd{\times} & 0 \\
       0 & 0 & \times
   \end{bmatrix}.
\end{equation}
%%%
%%%
Thus, the observed CKM matrix can be generated in this manner.

In the following section, we discuss the generation of lepton mixing and comment on the nature of neutrinos.   

%%%%%%%%%%%%%%%%%%%%%%%%%%%%%%%%%%%%%%%%%%%%%%%%%%%%%%%%%%%%%%%%%%%%%%%%%%%%%%%%%%%%%%%%%%%%%%%%%%%%%%%%%%%%%%%%%%%%%%%%%%%%%%%%% 
\section{Neutrino mass mechanism and Lepton mixing}\label{Sec:Lep_mix}
\begin{figure}[ht]
\begin{center}
\includegraphics[width=0.49\linewidth]{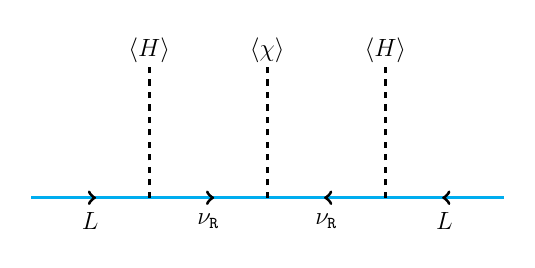}
\end{center}
\caption{Type-1 seesaw to generate Majorana neutrino masses. }
\label{FiG:Tree}
\end{figure}

%%%%%%%%%%%%%%%%%%%%%%%%%%%%%%%%%%%%%%%%%%%%%%%%%%%%%%%%%%%%%%%%%%%%%%%%%%%%%%%%%%%%%%%%%%%%%%%%%%%%%%%%%%%%%%%%%%%%%%%%%%%%%%%%%
In this section, we discuss the generation of lepton mixing and neutrino masses for models presented in Sec.~\ref{Sec:Quark_mixing}.
Within this framework, neutrinos can have either Dirac or Majorana nature. 
As noted earlier, both anomaly-cancellation solutions permit Dirac mass terms for neutrinos (see Sec. \ref{Sec:anomaly_cancellations}).
A Majorana nature can be realised through the UV completion of the generalized Weinberg operator $\overline{L^c} H H L \chi$ \cite{CentellesChulia:2018gwr}.
As discussed earlier, we consider an arbitrary number of BSM singlet scalars ($\chi_{i}$), focusing on the minimal configurations required to avoid massless Goldstone modes, and do not introduce any additional BSM fermions.
Consequently, the Weinberg operator can be realized via the type-I seesaw mechanism as illustrated in Fig.~\ref{FiG:Tree}
\footnote{Note that the generalized Weinberg operator $\overline{L^c} H H L \chi$ can also be UV-completed in many other ways; however, all such completions require the introduction of additional scalars, fermions, or both.}.
%%
%%%
%%%%
%%%%%%%%%%%%%%%%
%%%%%%%%%%%%%%%%
%%%%%%%%%%%%%%%%
%%%%%%%%%%%%%%%%
\subsection{Models with single Higgs doublet}\label{Subsec:Lmix_1hdm}
%%%%%%%%%%%%%%%%
%%%%%%%%%%%%%%%%
%%%%%%%%%%%%%%%%
Here we consider lepton mixing for the single Higgs case discussed in Table.~\ref{Tab:parametertable1}. 
The off-diagonal terms in the lepton
mass matrix can be written in two possible ways,
\begin{subequations}\label{Eq:Phi2_Charge_cross_term_Lepton}
\begin{align}
    &\overline{L^{3}} \tilde{\Phi} \nu_{\mathtt{R}}^{i}~( \text{or}~\overline{L^{i}} \Phi e_{\mathtt{R}}^{3}) ~ \longrightarrow X_{\Phi} = a + l - 2l_3 \label{Eq:Phi2_Charge_cross_term_Lepton1}\,,\\
    & \overline{L^{i}} \tilde{\Phi} \nu_{\mathtt{R}}^{3}~( \text{or}~\overline{L^{3}} \Phi e_{\mathtt{R}}^{i}) ~ \longrightarrow X_{\Phi} = a-l\label{Eq:Phi2_Charge_cross_term_Lepton2}\,. 
\end{align}    
\end{subequations}
%%

%%%%%%%%%%%%%%
By substituting $X_{\Phi}$ from Table~\ref{Tab:parametertable1} and solving, we find that Eq.~\eqref{Eq:Phi2_Charge_cross_term_Lepton1} and Eq.~\eqref{Eq:Phi2_Charge_cross_term_Lepton2} both yield the identical condition $l = l_3$.
%%%%%%%%%%%%%%
Furthermore, setting $l = l_3$ in Table~\ref{Tab:parametertable1} results in both quark and lepton sectors being flavor universal, with two free parameters remaining. This solution corresponds to the second solution discussed in Ref.~\cite{Prajapati:2024wuu}. 
In this case, the Weinberg operator can be UV-completed by  assigning the singlet scalar $\chi$ with $U(1)_X$ charge $-2a$, thereby generating Majorana neutrino masses.

When $l \neq l_3$, off-diagonal entries in the neutrino Dirac mass matrix cannot be generated with the existing scalar content. However, the type-I seesaw mechanism can still be realized by further extending the scalar sector.
In general, if $l \neq l_3$, UV-completing the Weinberg operator via Majorana mass terms for the right-handed neutrinos requires three additional singlet scalars with the following $U(1)_X$ charge assignments,
\begin{subequations}\label{Eq:Charge_Majorana_term_for_right_handed_fermion}
\begin{align} 
     & \overline{(\nu_{\mathtt{R}}^{i})^{c}} \nu_{\mathtt{R}}^{j} \chi_{1} ~\longrightarrow ~  X_{\chi_{1}} = -2(a+l-l_3) \,,\\
    & \overline{(\nu_{\mathtt{R}}^{i})^{c}} \nu_{\mathtt{R}}^{3} \chi_{2} ~\longrightarrow  ~ X_{\chi_{2}} = -(2a+l-l_3) \,,\\ 
     & \overline{(\nu_{\mathtt{R}}^{3})^c} \nu_{\mathtt{R}}^{3} \chi_{3} ~\longrightarrow ~  X_{\chi_{3}} = -2a \,.
\end{align}    
\end{subequations}
Note that, the $U(1)_X$ symmetry forbids cross terms between these scalars, resulting in three Goldstone bosons after SSB. One is absorbed by the $Z'$ boson to acquire its third polarization state, while the remaining two stay as physical massless states. This issue can be resolved by introducing an additional scalar $\chi_4$ with $U(1)_X$ charge $l-l_3$, which allows three cross terms among the scalars and removes all massless degrees of freedom. Hence in this way leptonic mixing can be generated without reducing the free parameter in the fermion charge assignment.
%%%%%%%%%%%%%%%%%%%%%%%%%%%%%%%%%%%%

%%%%%%%%%%%%%%%%%%%%%%%%%%%%
\subsection{Models with two Higgs doublet}\label{Subsec:Lmix_2hdm}
%%%%%%%%%%%%%%%%%%%%%%%%%%%%
\subsubsection{ Extensions of Scenario (I) }
In models employing two Higgs doublets $\Phi$ and $\varphi$, we begin with Scenario (I). One possible approach is to adopt the flavor universal quark charges from Table~\ref{Tab:parametertable1} while inducing lepton mixing via $\varphi$.
%%%%
The constraints on the $U(1)_X$ charge of $\varphi$ required to generate these off-diagonal terms in Dirac neutrino mass matrix are obtained from Eq.~\eqref{Eq:Phi2_Charge_cross_term_Lepton} by replacing $\Phi$ with $\varphi$.
%%%
Additionally, constructing a $U(1)_X$ invariant coupling between the two doublets requires a singlet scalar $\chi_0$ with $U(1)_X$ charge $l - l_3$, for both choices of $X_\varphi$ given in Eq.~\eqref{Eq:Phi2_Charge_cross_term_Lepton}.
%%%
The $U(1)_X$ charges of the particles in this scenario are presented in Table~\ref{Tab:parametertable3}.
%%%
%%%%%%
\begin{table}[ht]
\centering
%\captionsetup{justification=centering}
\renewcommand{\arraystretch}{1.5}
\setlength{\tabcolsep}{8pt} % control spacing between columns
\small
\begin{tabular}{|c c|c c|}
\hline
Fields &~~ $U(1)_X$ ~~& Fields & ~~$U(1)_X$~~ \\ 
\hline
$Q^{i}$ & $-\frac{1}{9}(2l+l_3)$ & $L^{i}$ & $l$ \\ 
$u_{\mathtt{R}}^{i}$ & $a-\frac{2}{9}(l+5l_3)$ & $e_{\mathtt{R}}^{i}$ & $l + l_{3} - a$ \\ 
$d_{\mathtt{R}}^{i}$ & $-a + \frac{2}{9}(4l_3 - l)$ & $\nu_{\mathtt{R}}^{i}$ & $a + l - l_{3}$ \\ 
\hline
$Q^{3}$ & $-\frac{1}{9}(2l+l_3)$ & $L^{3}$ & $l_{3}$ \\ 
$u_{\mathtt{R}}^{3}$ & $a-\frac{2}{9}(l+5l_3)$ & $e_{\mathtt{R}}^{3}$ & $2l_{3} - a$ \\ 
$d_{\mathtt{R}}^{3}$ & $-a + \frac{2}{9}(4l_3 - l)$ & $\nu_{\mathtt{R}}^{3}$ & $a$ \\ 
\hline
$\Phi$ & $a - l_{3}$ & $\chi_0$ & $l-l_3$ \\ 
$\varphi$ & $a - l$ or $a+l-2l_3$ & &  \\ 
\hline
\end{tabular}
\caption{Charges of SM and BSM particles under $U(1)_X$. Quark charges are flavor universal. Higgs doublet $\Phi$ generates the masses of all fermions and quark mixing, while $\varphi$ accounts for leptonic mixing.}
\label{Tab:parametertable3}
\end{table}
%%%%%%%%%%%%%%%%%%%%
Hence, $\chi_0$ alone, with charges shown in Table~\ref{Tab:parametertable3}, is sufficient in this case.
%%%%%%%%%%%%%%%%%%%%%
Interestingly, the $U(1)_X$ charges of $\chi_0$ and $\chi_4$ are identical. Hence, the additional scalar required to avoid Goldstone bosons in the type-I seesaw scenario naturally emerges in this case, see discussion below Eq. \eqref{Eq:Charge_Majorana_term_for_right_handed_fermion}. 

Another possibility within Scenario (I) is to adopt the charge assignment presented in Table~\ref{Table:parametertable2} and employ $\varphi$ to generate off-diagonal entries in both the quark and lepton mass matrices. In this case, one of the free parameters in the fermion charge assignments can be fixed to produce the desired off-diagonal terms in the lepton mass matrix.
With $X_{\varphi}$ fixed according to Eq.~\eqref{Eq:Phi2_Charge_cross_term1}, allowing the term $\overline{L^i}\tilde{\varphi}\nu_{\mathtt{R}}^3$ requires $y = a + 4(l - 4l_3)/9$, and to avoid Goldstone bosons, the corresponding $\chi_0$ must carry a $U(1)_X$ charge of $l_3 - l$. Alternatively, for the term $\overline{L^3}\tilde{\varphi}\nu_{\mathtt{R}}^j$, one obtains $y = a - 4(l_3 + 2l)/9$, with $\chi_0$ having a charge of $l - l_3$  \footnote{Choosing $X_\varphi$ as in Eq.~\eqref{Eq:Phi2_Charge_cross_term2} gives the same $y$ values, but interchanged relative to Eq.~\eqref{Eq:Phi2_Charge_cross_term1} for the two different choices.}.
In both cases, the same $\chi_0$ can generate the required cross term between the doublets, and its charge is again the same as $\chi_4$.  The $U(1)_X$ charges of particles for both cases are presented in Table \ref{Tab:parametertable4}.
%%%%%%%%%%%%
\begin{table}[ht]
\centering
\renewcommand{\arraystretch}{1.5}
\setlength{\tabcolsep}{6pt} % adjust column spacing
\small
\begin{minipage}{0.48\textwidth}
\centering
\begin{tabular}{|c c|c c|}
\hline
Fields & $U(1)_X$ & Fields & $U(1)_X$ \\ 
\hline
$Q^i$ & $-\frac{1}{9}(5l - 2 l_3)$ & $L^i$ & $l$ \\ 
$u_{\mathtt{R}}^i$ & $a - \frac{1}{9}(5l + 7 l_3)$ & $e_{\mathtt{R}}^{i}$ & $l + l_3 - a$ \\ 
$d_{\mathtt{R}}^{i}$ & $-a + \frac{1}{9}(11 l_3 - 5 l)$ & $\nu_{\mathtt{R}}^{i}$ & $a + l - l_3$ \\
\hline
$Q^3$ & $\frac{1}{9}(4l - 7 l_3)$ & $L^3$ & $l_3$ \\ 
$u_{\mathtt{R}}^{3}$ & $a + \frac{4}{9}(l - 4 l_3)$ & $e_{R_1}$ & $2 l_3 - a$ \\ 
$d_{\mathtt{R}}^{3}$ & $-a + \frac{1}{9}(4l + 2 l_3)$ & $\nu_{\mathtt{R}}^{3}$ & $a$ \\ 
 \hline
$\Phi$ & $a - l_3$ & $\chi_0$ & $l - l_3$ \\ 
$\varphi$ & $a + l - 2 l_3$ or $a - l$ & & \\ 
\hline
\end{tabular}
\end{minipage}
\hfill
\begin{minipage}{0.48\textwidth}
\centering
\begin{tabular}{|c c|c c|}
\hline
Fields & $U(1)_X$ & Fields & $U(1)_X$ \\ 
\hline
$Q^i$ & $\frac{1}{9}(l - 4 l_3)$ & $L^i$ & $l$ \\ 
$u_{\mathtt{R}}^{i}$ & $a + \frac{1}{9}(l - 13 l_3)$ & $e_{\mathtt{R}}^{i}$ & $l + l_3 - a$ \\ 
$d_{\mathtt{R}}^{i}$ & $ -a + \frac{1}{9}(l+5l_3)$ & $\nu_{\mathtt{R}}^{i}$ & $a + l - l_3$ \\ 
\hline
$Q^3$ & $-\frac{1}{9}(8l - 5 l_3)$ & $l_3$ & $l_3$ \\ 
$u_{\mathtt{R}}^{3}$ & $a - \frac{4}{9}(2l + l_3)$ & $e_{\mathtt{R}}^{3}$ & $2 l_3 - a$ \\ 
$d_{\mathtt{R}}^{3}$ & $-a + \frac{1}{9}(14l_3 - 8 l)$ & $\nu_{\mathtt{R}}^{3}$ & $a$ \\ 
\hline
$\Phi$ & $a - l_3$ & $\chi_0$ & $l - l_3$ \\ 
$\varphi$ &  $a - l$ or $a + l - 2 l_3$  & & \\ 
\hline
\end{tabular}
\end{minipage}
\caption{Charges of SM and BSM particles under $U(1)_X$. \textbf{Left:} this Table can be obtained from Table \ref{Table:parametertable2} by putting $y = a + 4(l - 4l_3)/9$.  
\textbf{Right:} this Table can be obtained from Table \ref{Table:parametertable2} by putting $y = a - 4(l_3 + 2l)/9$. For both models, the Higgs doublet $\Phi$ generates the masses of all fermions, while $\varphi$ induces mixing.} 
\label{Tab:parametertable4}
\end{table}
%%%%%%%%%%%%%%%%%%
A particularly interesting choice for the parameter is $a = 0$. With this, only two scalar singlets with $U(1)_X$ charges $l_3 - l$ and $2(l_3 - l)$ are needed to generate type-I seesaw, and no extra Goldstone bosons remain.
\subsubsection{Extensions of Scenario (II) }
We now consider models corresponding to Scenario~(II). In this case, two possible charge assignments that generate quark mixing are presented in Tables~\ref{Tab:parametertable5} and \ref{Tab:parametertable6}.
The $U(1)_X$ charges in Table~\ref{Tab:parametertable5} allow Dirac mass terms and mixing for both quarks and leptons, as shown in Eq.~\eqref{Eq:general_mass_marix_3}.
In this scenario, generating Majorana mass terms for the right-handed neutrinos requires three SM-singlet scalars $\chi_1$, $\chi_2$, and $\chi_3$ with $U(1)_X$ charges $-2(2l - x)$, $-2a$, and $-(a + 2l - x)$, respectively.
Sufficient cross terms involving these scalars and $\chi_0$ can be included to avoid massless Goldstone bosons, thereby enabling the type-I seesaw mechanism.
%%%
For the second charge assignment in Scenario~(II) (Table~\ref{Tab:parametertable6}), the corresponding mass matrices are given in Eq.~\eqref{Eq:general_mass_marix_4}.
Lepton mixing can again be realized by imposing a constraint on one of the fermionic charge parameters listed in Table~\ref{Tab:parametertable6}. A more elegant approach is to implement the type-I seesaw mechanism by introducing three additional singlet scalars with $U(1)_X$ charges $-4l + 2x$, $2x - 4(2l + l_3)/3$, and $2x - 2(5l + l_3)/3$.
In the presence of $\chi_0$, suitable cross terms among these scalars can be constructed to avoid massless Goldstone bosons.

The models presented in this section exhibit distinct features in their scalar content, fermion charge assignments, and resulting $Z'$ couplings, leading to diverse patterns in quark and lepton mixing as well as different possibilities for the nature of neutrino masses. These differences give rise to unique phenomenological signatures, such as the nature of vector and axial vector $Z'$ couplings to SM fermions,  B-physics anomalies in the angular observables and branching fractions to charged leptons, and neutrino interactions.
In the following section, we explore selected phenomenological implications of these models by analyzing a set of representative benchmark scenarios.
%%%

%%%%%%%%%%%%%%%%%%%%%%%%%%%%%%%%%%%%%%%
%%%%%%%%%%%%%%%%%%%%%%%%%%%%%%%%%%%%%%%
%%%%%%%%%%%%%%%%%%%%%%%%%%%%%%%%%%%%%%%
\section{Phenomenological Implications}\label{Sec:Phenomenological Implications}
%%%%%%%%%%%%%%%%%%%%%%%%%%%%%%%%%%%%%%%
%%%%%%%%%%%%%%%%%%%%%%%%%%%%%%%%%%%%%%%
%%%%%%%%%%%%%%%%%%%%%%%%%%%%%%%%%%%%%%%

Now we turn to the phenomenological implications of the models discussed in the previous section. The $U(1)_X$ charges of leptons and quarks can vary substantially depending on the specific anomaly cancellation solution and the chosen UV completion. Since these charges are flavor-dependent, each solution yields a distinct flavor structure in both the quark and lepton sectors. The resulting models are, in general, chiral, leading to both vector (V) and axial-vector (A) couplings of the $Z'$ to SM fermions. We begin by examining the cases that give rise to purely vector or purely axial-vector interactions.
%%%%%%%%%%%%%%%%%%%%%%%%%%%%%%%%%%%%%%%
%%%%%%%%%%%%%%%%%%%%%%%%%%%%%%%%%%%%%%%
%%%%%%%%%%%%%%%%%%%%%%%%%%%%%%%%%%%%%%%
\subsection{Pure Vector and Axial Vector Couplings}\label{subSec:Pure Vector and Axial Vector Couplings}
%%%%%%%%%%%%%%%%%%%%%%%%%%%%%%%%%%%%%%%
%%%%%%%%%%%%%%%%%%%%%%%%%%%%%%%%%%%%%%%
%%%%%%%%%%%%%%%%%%%%%%%%%%%%%%%%%%%%%%%
In this section, taking specific example model, we show how purely axial-vector or mixed vector-axial-vector coupling can be obtained. As discussed before, If $U(1)_X$ charge assignment of fermion is flavor universal, then the Higgs doublet $\Phi$ that acquires the  larger VEV and also generates the mass of this fermion through Yukawa interactions, the corresponding axial-vector current vanishes, as seen from Eq.~\eqref{Eq:VA_coup}.
%%%
As a consequence, flavor universal chiral models, such as linear combinations of hypercharge and $B-L$, effectively reduce to purely vector like theories in the light-$Z'$ regime.

In a two Higgs doublet setup with flavor specific fermion charges, the Higgs charge appearing in Eq.~\eqref{Eq:VA_coup} may differ from the Higgs charge required to generate mass via Yukawa coupling between fermions. Consequently, mixed vector-axial vector or purely axial-vector coupling can be achieved for lighter fermions in such a setup.
%%%
The anomaly cancellation solutions presented in Sec.~\ref{Subsec:SII_Anomaly} satisfy this requirement, where the Yukawa interactions that generate the masses of the first two generations are allowed through the Higgs doublet $\varphi$, while those for the third generation are allowed through $\Phi$. Consequently, in this scenario, the first and second generation quarks and leptons can have mixed vector-axial vector or purely axial-vector coupling. Purely axial-vector coupling can be achieved under the following condition,
%%%
\begin{equation}\label{EQ:vector_coupling_zero_condition2}
    X_{\Phi} = - \frac{C^{\psi}_{V(0)}}{ (T_{\psi_{\mathtt{L}}}^{3} - 2 Q_{\psi} \sin^{2} \theta_{W})}\,.
\end{equation}
%%%%%%
\begin{table}[ht]
\centering
%\captionsetup{justification=centering}
\renewcommand{\arraystretch}{1.5}
\setlength{\tabcolsep}{10pt} % balanced horizontal padding
\small
\begin{tabular}{|c c|c c|}
\hline
\textbf{Fields} & \textbf{$U(1)_X$} & \textbf{Fields} & \textbf{$U(1)_X$} \\ 
\hline
$Q^{i}$ & $\tfrac{s^{2}}{3}$ & $L^{i}$ & $-s^{2}$ \\ 
$u_{\mathtt{R}}^{i}$ & $-(1 - \tfrac{7}{3}s^{2})$ & $e_{\mathtt{R}}^{i}$ & $1 - 3s^{2}$ \\ 
$d_{\mathtt{R}}^{i}$ & $1 - \tfrac{5}{3}s^{2}$ & $\nu_{\mathtt{R}}^{i}$ & $-(1 - s^{2})$ \\ 
\hline
$Q^{3}$ & $2 - \tfrac{5}{3}s^{2}$ & $L^{3}$ & $-(6 - 5s^{2})$ \\ 
$u_{\mathtt{R}}^{3}$ &  $3 - \tfrac{5}{3}s^{2}$  & $e_{\mathtt{R}}^{3}$ & $-(7 - 5s^{2})$  \\ 
$d_{\mathtt{R}}^{3}$ & $1 - \tfrac{5}{3}s^{2}$ & $\nu_{\mathtt{R}}^{3}$ & $-5(1 - s^{2})$  \\ 
\hline
$\Phi$ & $1$ & $\chi_{0}$ & $2(1 - s^{2})$ \\ 
$\varphi$ & $-(1 - 2s^{2})$ &  &  \\ 
\hline
\end{tabular}
\caption{Charges of SM and BSM particles under $U(1)_X$ yielding pure axial-vector couplings for the first two generations in the light $Z'$ scenario.}
\label{Tab:parametertable7}
\end{table}
%%%%%%%%%%%%%%%%%%
As an illustrative case, consider the $U(1)_X$ charge assignment given in Table~\ref{Tab:parametertable6}. Using Eq.~\eqref{EQ:vector_coupling_zero_condition2}, the condition for the light quark and lepton couplings to be purely axial-vector is obtained as,
%%%
\begin{equation}
    l = - \sin^{2} \theta_{W},\,\, l_3 = - (6 - 5 \sin^{2} \theta_{W}), \,\, x = 1 - 3 \sin^{2} \theta_{W}\,.
\end{equation}
%%%

With this replacement the Table. \ref{Tab:parametertable7} shows the $U(1)_X$ charge assignments of particles. Here, $s$ denotes the sine of the Weinberg angle, $s \equiv \sin \theta_{W}$.  The resulting couplings for fermions with $Z'$ is listed in the Table. \ref{Tab:Coupling_for_zero_vector}.
\begin{table}[h]
\begin{center}
\begin{adjustbox}{max width=\textwidth}
\renewcommand{\arraystretch}{2}
\begin{tabular}{|c|c|c|c|c|c|c|c|}
\hline
\textbf{Fields} ($\psi$)& ~~~~~$g_{\psi}^{z'}$~~~~~ & \textbf{Fields} ($\psi$) & ~~~~~$g_{\psi}^{z'}$&~~~~~ \textbf{Fields} ($\psi$)& ~~~~~$g_{\psi}^{z'}$~~~~~ & \textbf{Fields} ($\psi$) & ~~~~~$g_{\psi}^{z'}$~~~~~\\  
\hline
$u_{\mathtt{L}}^{i}$ & $ g_{x} \cos^{2} \theta_{W} $ & $u_{\mathtt{R}}^{i}$ & $ - g_{x} \cos^{2} \theta_{W}$ & $e_{\mathtt{L}}^{i}$ & $ - g_{x} \cos^{2} \theta_{W}$ & $e_{\mathtt{R}}^{i}$ & $g_{x} \cos^{2} \theta_{W}$ \\
$d_{\mathtt{L}}^{i}$ & $ - g_{x} \cos^{2} \theta_{W}$ & $d_{\mathtt{R}}^{i}$ & $  g_{x} \cos^{2} \theta_{W}$ & $\nu_{\mathtt{L}}^{i}$ & $ g_{x} \cos^{2} \theta_{W}$ & $\nu_{\mathtt{R}}^{i}$ & $-g_{x} \cos^{2} \theta_{W}$ \\
\hline
$u_{\mathtt{L}}^{3}$ & $3 g_{x} \cos^{2} \theta_{W} $ & $u_{\mathtt{R}}^{3}$ & $ 3 g_{x} \cos^{2} \theta_{W}$& $e_{\mathtt{L}}^{3}$ & $-7 g_{x} \cos^{2} \theta_{W}$ & $e_{\mathtt{R}}^{3}$ & $-7g_{x} \cos^{2} \theta_{W}$\\
$d_{\mathtt{L}}^{3}$ & $  g_{x} \cos^{2} \theta_{W}$ & $d_{\mathtt{R}}^{3}$ & $  g_{x} \cos^{2} \theta_{W}$& $\nu_{\mathtt{L}}^{3}$ & $ - 5g_{x} \cos^{2} \theta_{W}$ & $\nu_{\mathtt{R}}^{3}$ & $-5g_{x} \cos^{2} \theta_{W}$ \\
\hline
\end{tabular}
\end{adjustbox}
\end{center}
\caption{Fermion couplings to the $Z'$ boson in the gauge basis. The first two generations have pure axial-vector couplings, while the third generation has pure vector couplings.}
\label{Tab:Coupling_for_zero_vector}
\end{table}
%%%%%%%%%%%

It is noteworthy that the couplings of the first two generations of quarks and leptons are purely axial-vector, whereas those of the third generation are purely vectorial. However, these couplings are expressed in the fermion gauge basis. Since the model is generation-specific, rotation to the mass basis may, in general, induce vector components. Nonetheless, appropriate choices of the rotation matrices can render the vector couplings negligible or exactly vanishing.
%%%%%%%%%%%%%%%%%%%%%%
%%%%%%%%%%%%%%%%%%%%%%
\subsection{Evading Neutrino Scattering Bounds in the Light $Z'$ Regime}\label{subsec:Neutrino_couling}
%%%%%%%%%%%%%%%%%%%%%%
%%%%%%%%%%%%%%%%%%%%%%
Low-energy neutrino experiments, such as neutrino-electron scattering and coherent elastic neutrino-nucleus scattering, impose some of the most stringent constraints on the gauge coupling $g_x$ of a new light vector or axial-vector mediator \cite{Majumdar:2024dms,DeRomeri:2024dbv}. 
%%%
These bounds can be significantly relaxed if the $U(1)_X$ charges of neutrinos are taken to be zero or small. However, in most models, the anomaly cancellations conditions along with the fact that neutrinos are part of the lepton doublets, implies that  this flexibility is absent. Thus, the parameter space with a light $Z'$ is therefore severely constrained by null BSM search results from neutrino scattering experiments.
%%%

Recall that, in our framework, when the $\rho$-parameter constraints are satisfied and $M_{Z'}^{2}/M_{Z}^{2} \ll 1$, the mixing angle $\alpha$ becomes effectively independent of $M_{Z'}$.
The resulting couplings in the gauge basis are presented in Eq.~\eqref{EQ:Zp Couplings} and are given by
\begin{equation}\label{EQ:Zp Couplings11}
g_{\psi_{_{\mathtt{L}}}}^{z'} \simeq  \left[2 X_{\Phi} \Big(T_{\psi_{\mathtt{L}}}^{3} - Q_{\psi} \sin^{2} \theta_{W} \Big)+ X_{\psi_{_{\mathtt{L}}}} \right]g_x.
\end{equation}
For SM neutrino, $T_{\nu_{\mathtt{L}}}^{3}=1/2$ and its electric charge is zero, $Q_{\psi}=0$. 
Hence Eq. \eqref{EQ:Zp Couplings11} reduces to, $g_{\nu_{\mathtt{L}}}^{z'} \simeq (X_{\Phi} + X_{L})\,g_x$.
Therefore,  the neutrino-$Z'$ coupling vanishes when the Higgs doublet with the larger VEV carries a $U(1)_X$ charge opposite to that of the lepton doublet.

%%%
As an example model, consider the one presented in Table~\ref{Tab:parametertable5}. The $U(1)_X$ charge of the lepton doublets is $l$, while the Higgs carries charge $a-l$. In this case, the resulting coupling of SM neutrinos in the fermionic gauge basis is ``$a\, g_x$''. Since $a$ is a free parameter, it can be adjusted to relax the constraints on the gauge coupling $g_x$. Infact one is free to even take the limit $a \to 0$, in which case the SM neutrino completely decouples from the $Z'$.
%%%%
%%%%
Hence, the  model is free from constraints arising from neutrino-electron (nucleus) scattering,
which is otherwise a major constraint in low $M_{Z'}$ limit.
%%%%%%%%%%%%%%%%%%%%%%
%%%%%%%%%%%%%%%%%%%%%%
%%%%%%%%%%%%%%%%%%%%%%
\subsection{B anomalies}
%%%%%%%%%%%%%%%%%%%%%%
%%%%%%%%%%%%%%%%%%%%%%
%%%%%%%%%%%%%%%%%%%%%%
The recent LHCb measurements show that the lepton flavor universality ratios $R_K$ and $R_{K^*}$ are compatible with the SM~\cite{LHCb:2022qnv,LHCb:2022vje}. Consequently, attention has refocused on the persistent anomalies in the angular observables and branching ratios of $b \to s$ transitions~\cite{LHCb:2013ghj,LHCb:2014cxe,LHCb:2015tgy,LHCb:2015wdu,LHCb:2015svh,LHCb:2020lmf,LHCb:2020gog,LHCb:2021xxq,LHCb:2021zwz,CMS:2024atz}.
%%%
%%%
The anomaly cancellation solutions presented in Sec.~\ref{Sec:anomaly_cancellations} feature flavor specific $U(1)_X$ charge assignments, and can therefore induce tree-level FCNCs mediated by the $Z'$. At the same time, the first two generations of SM fermions carry identical $U(1)_X$ charges, consistent with the lepton flavor universality observed between electrons and muons in LHCb measurements. Consequently, this framework can be employed to address the anomalies in the angular observables and branching ratios of $b \to s$ transitions.
%%%
%%%
To highlight this, we take the scalars to be heavy and/or the off-diagonal Yukawa entries to be small, and hence we focus only on the $Z'$-mediated FCNCs while neglecting scalar-mediated FCNCs.
%%%
%%%

%%%%%%%%%%%%%%%%%%%%%%
The effective Hamiltonian responsible for $b \to s l^{+} l^{-}$ transition  can be written as,
\begin{equation}
    \mathcal{H}_{\text{eff}}^{bs\ell \ell} = \frac{-4 G_{f}}{\sqrt{2}} \frac{\alpha_{\text{EM}}}{4 \pi} V_{tb}V_{ts}^{*} \sum_{i =9/10} \left(  C_{i}^{\ell \ell}\mathcal{O}_{i}^{\ell \ell} + C_{i}^{\ell \ell '}\mathcal{O}_{i}^{\ell \ell '} \right)\,.
\end{equation}
Here, $G_f$ and $\alpha_{\text{EM}}$ denote the Fermi constant and the electromagnetic fine-structure constant, respectively. $V_{tb(s)}$ represents the relevant CKM matrix element, and $\ell = e, \mu$. The quantities $C_{i}^{\ell \ell\,(\prime)}$ are the effective Wilson coefficients evaluated at the bottom-quark mass scale. The corresponding effective operators are given by,
\begin{subequations}\label{Wilson_operator}
\begin{align}
    & \mathcal{O}_{9}^{\ell \ell} = (\Bar{s} \gamma_{\mu}P_{\mathtt{L}}b)(\Bar{\ell} \gamma^{\mu} \ell),\,~~~~~~\mathcal{O}_{9}^{\ell \ell '} = (\Bar{s} \gamma_{\mu}P_{\mathtt{R}}b)(\Bar{\ell} \gamma^{\mu} \ell)\,,\\
    &  \mathcal{O}_{10}^{\ell \ell} = (\Bar{s} \gamma_{\mu}P_{\mathtt{L}}b)(\Bar{\ell} \gamma^{\mu} \gamma_{5} \ell),\,~~~~~~\mathcal{O}_{10}^{\ell \ell'} = (\Bar{s} \gamma_{\mu}P_{\mathtt{R}}b)(\Bar{\ell} \gamma^{\mu} \gamma_{5} \ell)\,. 
\end{align}    
\end{subequations}
The Wilson coefficients are written as $C_{i}^{\,ll(\prime)} = (C_{i}^{\ell \ell(\prime)})_{\text{SM}} + \Delta C_{i}^{\,ll(\prime)}$, where $(C_{i}^{\ell \ell(\prime)})_{\text{SM}}$ denotes the SM contribution and $\Delta C_{i}^{\,ll(\prime)}$ captures the new physics effects.
In general, the Lagrangian density corresponding to these $Z'$ mediated interactions can be written as,
\begin{equation}\label{Eq:FCNC_Lag_Zp_Int}
    - \mathcal{L} = \left( g_{\mathtt{L}}^{\ell \ell} \overline{\ell}\gamma^{\mu} P_{\mathtt{L}} \ell + g_{\mathtt{R}}^{\ell \ell} \overline{\ell}\gamma^{\mu} P_{\mathtt{R}} \ell + g_{\mathtt{L}}^{bs} \Bar{s}\gamma^{\mu} P_{\mathtt{L}}b + g_{\mathtt{R}}^{bs} \Bar{s}\gamma^{\mu} P_{\mathtt{R}}b  \right)Z_{\mu}^{'} + \text{h.c.}\,. 
\end{equation}
Here, $g_{\mathtt{L}(\mathtt{R})}^{\ell\ell}$ and $g_{\mathtt{L}(\mathtt{R})}^{bs}$ denote the left- (right-) handed couplings of the $Z'$ to charged leptons and to the bottom-strange quark current, respectively. The corresponding new physics contribution to Wilson coefficients at tree level are given as,
%%%%
\begin{subequations}\label{Wilson_Coeff}
\begin{align}
    & \Delta C_{9}^{\ell \ell} = - \frac{\pi}{\sqrt{2} G_{f} \alpha_{\text{EM}}} \frac{1}{V_{tb}V_{ts}^{*}} \frac{g_{\mathtt{L}}^{bs} (g_{\mathtt{L}}^{\ell \ell} + g_{\mathtt{R}}^{\ell \ell})}{M_{Z'}^{2}}\,, ~~~ \Delta C_{9}^{\ell \ell '} = - \frac{\pi}{\sqrt{2} G_{f} \alpha_{\text{EM}}} \frac{1}{V_{tb}V_{ts}^{*}} \frac{g_{\mathtt{R}}^{bs} (g_{\mathtt{L}}^{\ell \ell} + g_{\mathtt{R}}^{\ell \ell})}{M_{Z'}^{2}}\\
    & \Delta C_{10}^{\ell \ell} = \frac{\pi}{\sqrt{2} G_{f} \alpha_{\text{EM}}} \frac{1}{V_{tb}V_{ts}^{*}} \frac{g_{\mathtt{L}}^{bs} (g_{\mathtt{L}}^{\ell \ell} - g_{\mathtt{R}}^{\ell \ell})}{M_{Z'}^{2}}\,,~~ \Delta C_{10}^{\ell \ell '} = \frac{\pi}{\sqrt{2} G_{f} \alpha_{\text{EM}}} \frac{1}{V_{tb}V_{ts}^{*}} \frac{g_{\mathtt{R}}^{bs} (g_{\mathtt{L}}^{\ell \ell} - g_{\mathtt{R}}^{\ell \ell})}{M_{Z'}^{2}}\,. 
\end{align}    
\end{subequations}
%%%%%%%%%%%%%%%%%%

The form of the Lagrangian  presented in Eq.~\eqref{Eq:FCNC_Lag_Zp_Int}, together with the associated Wilson coefficients in Eq.~\eqref{Wilson_Coeff}, can be derived from the models discussed in Secs.~\ref{Sec:Quark_mixing}--\ref{Sec:Lep_mix} by performing the rotation of fermion fields from the gauge basis to the mass basis.
This rotation is performed by diagonalizing the fermion mass matrices via bi-unitary transformations.
The corresponding field rotations are given by,
%%%%%%%%%%
\begin{equation}
    \psi_{\mathtt{L}}^{m} = U_{\psi_{_\mathtt{L}}}^{\dagger}\psi_{\mathtt{L}},\,\,  \psi_{\mathtt{R}}^{m} = U_{\psi_{_\mathtt{R}}}^{\dagger}\psi_{\mathtt{R}}.
\end{equation}
%%%%%%%%%%
These rotations induce fermion mixing and generate FCNCs in both the quark and lepton sectors. To prevent FCNCs in the charged-lepton sector, we adopt the basis in which the charged-lepton mass matrix is diagonal, corresponding to $U_{e_L} = U_{e_R} = \mathtt{1}$.
%%%%
In the quark sector, we begin with the charged-current interactions. Here, the CKM matrix arises from the unitary transformations that diagonalize the up-type and down-type quark mass matrices,
%%%%%%%%%%%
\begin{equation}
    -\mathscr{L}_{WCC}^{q}= \frac{g}{\sqrt{2}} \left[  \overline{d_{_\mathtt{L}}^{m}} \gamma^{\mu} V_{CKM}^{\dagger} u_{_\mathtt{L}}^{m}W_{\mu}^{-} + \overline{u_{_\mathtt{L}}^{m}} \gamma^{\mu} V_{CKM} d_{_\mathtt{L}}^{m}W_{\mu}^{+}  \right].
\end{equation}
%%%%%%%%%%%%%
Where, $d_{_\mathtt{L}}^{m} = ( d_{\mathtt{L}}, s_{\mathtt{L}}, b_{\mathtt{L}})^{T} $,  $u_{_\mathtt{L}}^{m} = (u_{\mathtt{L}},c_{\mathtt{L}},t_{\mathtt{L}})^{T}$, and $V_{CKM} =U_{u_{_\mathtt{L}}}^{\dagger}U_{d_{_\mathtt{L}}} $. 
We adopt the standard parametrization in which the up quark mass matrix is taken to be diagonal, so that $U_{u_L} = U_{u_R} = \mathtt{1}$.
In this basis, the CKM matrix is  
%%%%%%%%%%%%%%%%
\begin{eqnarray}
    V_{CKM} = U_{d_{_\mathtt{L}}} = \begin{pmatrix}
        V_{ud} & V_{us} & V_{ub} \\
        V_{cd} & V_{cs} & V_{cb} \\
        V_{td} & V_{ts} & V_{tb}
    \end{pmatrix}\,.
\end{eqnarray}
%%%%%%%%%%%%%%%%

Now, we will focus on the neutral current sector of quarks mediated by both $Z$ and $Z'$ bosons. The new symmetry will give us new interactions in this sector. 
In the heavy $Z'$ scenario, the mixing angle $\alpha$ is suppressed, so the neutral current interactions mediated by the $Z$ boson are effectively identical to those in the SM within this framework.
The new interactions  mediated by $Z'$ give rise to FCNC, as shown below,
%%%%%%%%%%%%%%%%
\begin{align}
    J_{\mu}^{Z'} = \overline{d_{_\mathtt{L}}^{m}} \gamma_{\mu} U_{d_{_\mathtt{L}}}^{\dagger} \begin{pmatrix}
        g_{d_{{\mathtt{L}}}^{1}}^{z'} & 0 & 0 \\
        0 & g_{d_{{\mathtt{L}}}^{2}}^{z'} & 0 \\
        0 & 0 & g_{d_{_{\mathtt{L}}}^{3}}^{z'}  
    \end{pmatrix}
    U_{d_{_\mathtt{L}}} d_{\mathtt{L}}^{m} + \overline{d_{_{_\mathtt{R}}}^{m}} \gamma_{\mu} U_{d_{_\mathtt{R}}}^{\dagger}
    \begin{pmatrix}
        g_{d_{{\mathtt{R}}}^{1}}^{z'} & 0 & 0 \\
        0 & g_{d_{{\mathtt{R}}}^{2}}^{z'} & 0 \\
        0 & 0 & g_{d_{{\mathtt{R}}}^{3}}^{z'}  
    \end{pmatrix}
    U_{d_{_\mathtt{R}}}d_{\mathtt{R}}\,.
\end{align}
%%%
Here, $g_{d_{\mathtt{L(R)}}^{\,i}}^{Z'}$ denotes the $Z'$ coupling to the $i$-th generation down-type quark in its gauge basis, and could be obtained from Eq.~\eqref{SM Couplings} with replacement $\alpha \to - \alpha$.
%%%
For simplicity, we take $U_{d_{\mathtt{R}}} = \mathtt{1}$, which makes the right-handed down quark sector flavor diagonal. As a result, the coupling $g_{\mathtt{R}}^{bs}$ vanishes, and therefore no right-handed flavor-changing contribution is generated. 
%%%%
Consequently, the Wilson coefficients $\Delta C_{9(10)}^{\ell \ell'}$ vanish, and only $\Delta C_{9(10)}^{\ell \ell}$ are generated. The resulting current can be written as,
%%%%
\begin{align}
  J_{\mu}^{Z'} =  \begin{pmatrix}
          \overline{d_{L}} &  \overline{s_{L}} & \overline{b_{L}}
    \end{pmatrix} \gamma_{\mu}  V_{Z'}
    \begin{pmatrix}
        d_{\mathtt{L}} \\
        s_{\mathtt{L}} \\
        b_{\mathtt{L}}
    \end{pmatrix} + \begin{pmatrix}
          \overline{d_{R}} &  \overline{s_{R}} & \overline{b_{R}}
    \end{pmatrix} \gamma_{\mu}  \begin{pmatrix}
        g_{d_{_{\mathtt{R}}}}^{z'} & 0 & 0 \\
        0 & g_{s_{_{\mathtt{R}}}}^{z'} & 0 \\
        0 & 0 & g_{b_{_{\mathtt{R}}}}^{z'}  
    \end{pmatrix}
    \begin{pmatrix}
        d_{\mathtt{R}} \\
        s_{\mathtt{R}} \\
        b_{\mathtt{R}}
    \end{pmatrix}
\end{align}
%%%%%%%%%%%%%%%%
where,
%%%%%%%%%%%%%%%%
\begin{align}\label{Eq:Z'_FCNC_coupling_Mat}
V_{Z'} = V_{CKM}^{\dagger} \begin{pmatrix}
    g_{d_{{\mathtt{L}}}^{1}}^{z'} & 0 & 0 \\
        0 & g_{d_{{\mathtt{L}}}^{2}}^{z'} & 0 \\
        0 & 0 & g_{d_{_{\mathtt{L}}}^{3}}^{z'}  
    \end{pmatrix} V_{CKM}.
\end{align}
%%%%%%%%%%%%%%%%

From Eq.~\eqref{Eq:Z'_FCNC_coupling_Mat}, we obtain the $Z'$ mediated  FCNCs in the quark sector.
This framework can be employed to address the long standing discrepancies observed in the angular observables and branching ratios of $b \to s$ transitions. 
As an illustrative example, we consider the first solution listed in Table~\ref{Tab:parametertable4}. To simplify the analysis, we set $a = l_3$, which makes the Higgs charge $X_{\Phi}$ equal to zero, thereby reducing the model to a generation-specific vector scenario. The resulting charges are shown in the Table \ref{Tab:parametertable8}.
%%%%%%%%%%%%
\begin{table}[ht]
\centering
%\captionsetup{justification=centering}
\renewcommand{\arraystretch}{1.5}
\setlength{\tabcolsep}{8pt} % control spacing between columns
\small
\begin{tabular}{|c c|c c|}
\hline
Fields & $U(1)_X$ & Fields & $U(1)_X$ \\ 
\hline
$Q^i$ & $-\frac{1}{9}(5l - 2 l_3)$ & $L^i$ & $l$ \\ 
$u_{\mathtt{R}}^i$ & $-\frac{1}{9}(5l - 2 l_3)$ & $e_{\mathtt{R}}^{i}$ & $l$ \\ 
$d_{\mathtt{R}}^{i}$ & $-\frac{1}{9}(5l - 2 l_3)$ & $\nu_{\mathtt{R}}^{i}$ & $l$ \\ 
\hline
$Q^3$ & $\frac{1}{9}(4l - 7 l_3)$ & $L^3$ & $l_3$ \\ 
$u_{\mathtt{R}}^{3}$ & $\frac{1}{9}(4l - 7 l_3)$ & $e_{\mathtt{R}}^{3}$ & $l_3$ \\ 
$d_{\mathtt{R}}^{3}$ & $\frac{1}{9}(4l - 7 l_3)$ & $\nu_{\mathtt{R}}^{3}$ & $l_3$ \\ 
\hline
$\Phi$ & $0$ & $\chi_0$ & $l - l_3$ \\ 
$\varphi$ & $l_3 - l$  & & \\ 
\hline
\end{tabular}
\caption{Charges of SM and BSM particles under $U(1)_X$. The Higgs doublet $\Phi$ generates the masses of all fermions, while $\varphi$ induces mixing. The fermions carry vector like charges, $\Delta C_{10}^{\ell \ell}$ vanishes, and only $\Delta C_{9}^{\ell \ell}$ remains non-zero}
\label{Tab:parametertable8}
\end{table}
Since the Higgs charge $X_{\Phi}$ is zero, the mixing angle $\alpha$ also vanishes, and the $Z'$ couplings to left (right) handed leptons reduce to $g_x X_{\psi_{\mathtt{L(R)}}}$\footnote{In this case $Z$ boson couplings also reduces to its SM values.}. 
In this case, as the leptons carry vector like charges, as a result $\Delta C_{10}^{\ell \ell}$ vanishes, and only $\Delta C_{9}^{\ell \ell}$ remains non-zero, taking identical values for electrons and muons, i.e., $\Delta C_{9}^{e e} = \Delta C_{9}^{\mu \mu}$.

A global two-dimensional fit to all $b \to s$ observables, including both angular observables and branching ratios, yields the following best-fit values for $\Delta C_{9}^{\ell \ell}$ \cite{Hurth:2023jwr,Athron:2023hmz,Hurth:2025vfx},
\begin{equation}\label{Eq:Exp_C9}
    \{\Delta C_{9}^{\mu \mu}, \Delta C_{9}^{e e} \} = \{ -0.97 \pm 0.14, -0.77 \pm 0.18 \}
\end{equation}
For the numerical analysis, we fix the particle charges by choosing the benchmark values $l_3 = -2$ and $l = -3$. Under this choice, the matrix $V_{Z'}$ defined in Eq. \eqref{Eq:Z'_FCNC_coupling_Mat} takes the following numerical form,
%%%%%%%%%%%%%%%%%%%%%
\begin{align}\label{Eq:Vz'value}
|V_{Z'}| \sim  \begin{pmatrix}
    1.2 & ~~8\times 10^{-5}~~ &~~ 8.5 \times 10^{-3}~~ \\
        8\times 10^{-5} & 1.2 & 4 \times 10^{-2} \\
        8.5 \times 10^{-3} & 4 \times 10^{-2} & 0.2  
    \end{pmatrix} g_x.
\end{align}
%%%%%%%%%%%%%
Hence, mixing between the first and second generations is strongly suppressed compared to the mixings involving the third generation. The BSM contribution to the $b \to s$ transition arises from $g_{\mathtt{L}}^{bs} \sim 0.04 g_x $. 
%%%%
This coupling also induces a tree-level contribution to $B_s - \overline{B_s}$ mixing, which is loop-suppressed in the SM \cite{DiLuzio:2019jyq,DiLuzio:2017fdq,Albrecht:2024oyn}.
%%%%
The remaining flavor-changing elements in Eq.~\eqref{Eq:Vz'value} similarly generate tree-level contributions to other neutral meson mixings such as $B_d - \overline{B_d}$ and $K^{0} - \overline{K^{0}}$ \cite{Charles:2013aka,DiLuzio:2019jyq,Davighi:2021oel,Albrecht:2024oyn}. These processes collectively impose important constraints on the $g_x/M_{Z'}$ parameter space.
%%%%
%%
\begin{figure}[ht]
\begin{center}
\includegraphics[width=0.49\linewidth]{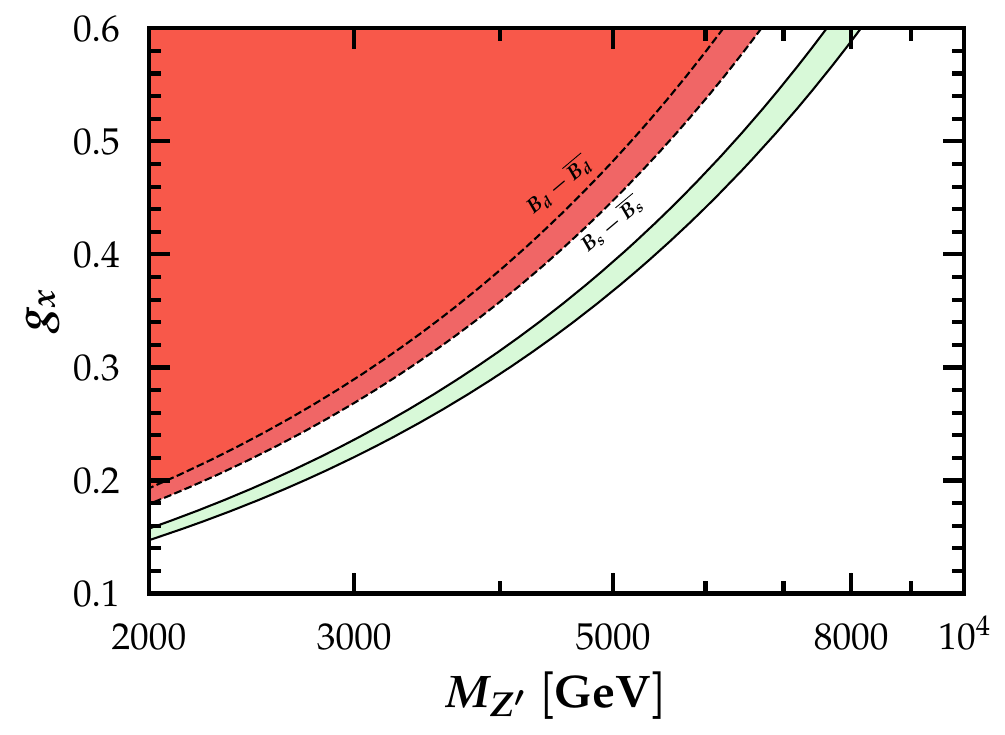}
\end{center}
\caption{ Mass of the $Z'$ versus its gauge coupling $g_x$. The green band shows the region consistent with the current global two-dimensional fit to all $b \to s$ observables. The parameter space excluded by $B_s - \overline{B_s}$ and $B_d - \overline{B_d}$ mixing is shown in red.}
\label{FiG:FCNC}
\end{figure}
In Figure~\ref{FiG:FCNC}, we show the parameter space in the ($M_{Z'}-g_x$) plane. The region consistent with the values of $\Delta C_{9}^{e e}$ and $\Delta C_{9}^{\mu \mu}$ from the current global two-dimensional fit to all $b \to s$ observables, as given in Eq.~\eqref{Eq:Exp_C9}, is highlighted in green. 
The parameter space excluded by $B_s - \overline{B_s}$ and $B_d - \overline{B_d}$ mixing is displayed in red. The constraints from $K^{0} - \overline{K^{0}}$ mixing are comparatively weak and therefore not shown.
As illustrated in Fig.~\ref{FiG:FCNC}, the green region, which corresponds to a global two-dimensional fit to all $b \to s$ observables (angular observables and branching ratios), is also consistent with the other FCNC constraints. 
Hence, this example model not only accommodates the current experimental indications of $b \to s$ anomalies but also remains compatible with the most stringent limits on flavor-changing interactions arising from neutral meson mixing. 
Additional constraints arise from collider searches for a $Z'$ in dilepton resonance channels, since both quarks and leptons carry $U(1)_X$ charges \cite{Prajapati:2024wuu}. These bounds are most stringent up to about 6 TeV; beyond this energy scale, resonant $Z'$ production becomes kinematically inaccessible at current colliders \cite{ATLAS:2019erb,CMS:2021ctt}. As shown in Fig.~\ref{FiG:FCNC}, for $M_{Z'} \gtrsim ~6$ TeV, a viable region of parameter space remains that satisfies all other constraints and is consistent with global fits to the $b \to s$ observables for reasonable values of $g_x$. We therefore do not include these collider limits in our analysis for this example model and leave a detailed study for future work.
%%%%%%%%%%%%%%%
%%%%%%%%%%%%%%%
%%%%%%%%%%%%%%%
\section{Conclusion}\label{Sec:Conclusion}
%%%%%%%%%%%%%%%
%%%%%%%%%%%%%%%
%%%%%%%%%%%%%%%
In this work, we have developed a class of flavor specific chiral $U(1)_X$ gauge extensions of the SM that provide new anomaly-free charge assignments capable of generating phenomenologically relevant pure vector, pure axial-vector, and mixed vector-axial interactions. By assigning identical $U(1)_X$ charges to the first two fermion generations and allowing the third generation to transform differently, the framework remains consistent with current measurements of lepton flavor universality while enabling nontrivial flavor structures required for B-physics anomalies and neutrino non-standard interactions.

Gauge and mixed anomaly cancellation is ensured through the inclusion of three right-handed neutrinos, and we have systematically classified all viable solutions consistent with fermion mass generation using at most two Higgs doublets and a minimal set of SM-singlet scalars. For each class of solutions, we have constructed UV-complete realisations illustrating how distinct flavor patterns and coupling structures emerge.

The benchmark models we presented span the range of possible charge assignments under which the quark and lepton $U(1)_X$ charges can differ substantially, leading to distinct phenomenological signatures. We analysed both the light and heavy $Z'$ scenarios and discussed the electroweak constraints from the $\rho$ parameter, as well as the resulting coupling structure of the $Z'$ in each case. 
Notably, we have demonstrated that mixed vector-axial vector or pure axial-vector couplings of the light $Z'$ to SM fermions, which are difficult to obtain in conventional vector or chiral models, can arise naturally within our framework.
Finally, using a representative benchmark model, we showed that this framework can address the current global two-dimensional fits to $b \to s$ observables while remaining compatible with constraints from other flavor-changing neutral-current processes such as $B_s - \overline{B_s}$, $B_d - \overline{B_d}$, and $K^0 - \overline{K^0}$ mixing.

In conclusion, this framework offers multiple possibilities and considerable freedom in assigning fermionic $U(1)_X$ charges for constructing flavor specific models capable of addressing various experimental anomalies. It also provides a theoretically well-motivated setting for obtaining phenomenologically interesting pure axial-vector and mixed vector–axial-vector couplings.

\appendix

%%%%%%%%%

\bibliographystyle{utphys}
\bibliography{bibliography}

\end{document}